\documentclass[a4paper,11pt]{article}

\usepackage[T1]{fontenc}
\usepackage[utf8]{inputenc}

\usepackage{geometry}
\usepackage{xcolor}
\usepackage{amsmath, array, amssymb, amsfonts}
\usepackage{mathtools}
\usepackage{amsthm}
\usepackage{dsfont}

\usepackage[french, english]{babel}
\usepackage[all]{xy}

\usepackage{fourier-orns}
\usepackage{textcomp}
\usepackage{lmodern}

\usepackage{array,booktabs}
\usepackage{caption}
\usepackage[hidelinks]{hyperref}

\DeclareMathAlphabet{\mathpzc}{OT1}{pzc}{m}{it}

\newcommand{\bs}[1]{\boldsymbol{#1}}
\newcommand{\defeq}{\vcentcolon=}

\renewcommand\P{\mathcal{P}}
\newcommand\M{\mathcal{M}}

\newcommand\doubleR{\mathbb{R}}

\renewcommand\1{\textbf{1}}

\renewcommand\H{\mathcal{H}}
\newcommand\A{\mathcal{A}}

\newcommand\U{\mathcal{U}}
\newcommand\SO{\mathcal{SO}}

\newcommand\K{\mathcal{K}}

\newcommand\D{\mathcal{D}}

\newcommand\rarrow{\rightarrow}

\newcommand\Rmk{\textit{Remark}}

\newcommand\LieG{\mathfrak{g}}
\newcommand\LieH{\mathfrak{h}}
\newcommand\Liep{\mathfrak{p}}

\newcommand\so{\mathfrak{so}}
\newcommand\co{\mathfrak{co}}
\newcommand\gl{\mathfrak{gl}}
\renewcommand\u{\mathfrak{u}}
\newcommand\LieK{\mathfrak{k}}

\renewcommand\tilde{\widetilde}
\newcommand\h{\widehat}
\renewcommand\b{\bar }

\renewcommand\d{\partial}

\renewcommand\-{^{-1}}
\newcommand\Ad{\text{Ad}}

\renewcommand\1{\mathds{1}}

\newtheorem{thm}{Theorem}
\newtheorem{lem}[thm]{Lemma}
\newtheorem{cor}[thm]{Corollary}

\theoremstyle{definition}

\hypersetup{
pdfauthor={J. François, S. Lazzarini, T. Masson},
pdftitle={Residual Weyl symmetry out of conformal geometry and its BRS structure.},
pdfsubject={},
pdfcreator={pdflatex},
pdfproducer={pdflatex},
pdfkeywords={}
}

\begin{document}

\begin{titlepage}

\title{Residual Weyl symmetry out of conformal geometry\\ and its BRS structure.}
\author{J. François, S. Lazzarini and T. Masson}
\date{}
\maketitle

\begin{center}
\vskip -5mm
Centre de Physique Théorique,\\
{\small Aix Marseille Université \& Université de Toulon \& CNRS UMR 7332,\\
Case 907, 13288 Marseille, France.}
\end{center}

\begin{abstract}
The conformal structure of second order in $m$-dimensions together with the so-called (normal) conformal Cartan connection, is considered as a framework for gauge theories.
The dressing field scheme presented in a previous work amounts to a decoupling of both the inversion 
and the Lorentz symmetries such that the residual gauge symmetry is the Weyl symmetry. On the one hand, it provides straightforwardly the \emph{Riemannian parametrization} of the normal conformal Cartan connection and its curvature. On the other hand, it also provides the finite transformation laws under the Weyl rescaling of the various geometric objects involved.
Subsequently, the dressing field method is shown to fit the BRS differential algebra treatment of infinitesimal gauge symmetry. The dressed ghost field encoding the residual Weyl symmetry is presented. The related so-called algebraic connection supplies relevant combinations found in the literature in the algebraic study of the Weyl anomaly.
\end{abstract}

\end{titlepage}

\newpage


\tableofcontents

\section{Introduction}

The use of Cartan geometry in physics is ideally suited for the formulation of space-time transformations as gauge theories as well as for the study of geodesics, see for instance~\cite{Harnad-Pettitt1,Friedrich:1995vb}. General Relativity (GR) can be incorporated into the scheme of gauge theories under its Einstein-Cartan formulation~\cite{Utiyama:1956sy,Kibble:1961ba}, see also~\cite{Hehl:1976kj} and references therein. Relatively recent work~\cite{Wise:2006sm,Lazzarini-Tidei,Wise:2009fu} shows a renewed interest in the use of Cartan connections for gravity theories.

In particular, conformal gravity is based on the conformal group, which contains the Poincaré group, the Weyl symmetries (rescaling of a metric) and the special conformal transformations (inversions). Since the latter cannot be globally defined, gauging the inversions as local transformations is recommended. The mathematics of the second order conformal structure is well understood, see for instance~\cite{Ogiue,Kobayashi,Bas91,AM95,CSS1997b,Sharpe}, and, as a framework for gauge theories, it deserves to be studied for itself~\cite{Harnad-Pettitt2,Harnad-Pettitt3,Coleman:1980ur}.

\smallskip
Regarding gauge symmetries in field theories, in~\cite{GaugeInv,JF_PhD} was described a systematic procedure to reduce gauge symmetries. The advocated scheme turns out to be a redefinition of the fields variables contained in the gauge theory at hand. It is grounded on the identification, among the fields of the theory, of what we called a \emph{dressing field}. 

The current paper can be considered as a continuation of~\cite{GaugeInv} since we provide a new example of such a reduction of symmetries.
We consider the second order conformal structure as a framework for gravity gauge theories. Using the dressing field method, we locally reduce the second order conformal structure to its Weyl subbundle. 
Then, we show that the dressing field method is compatible with the usual Becchi-Rouet-Stora (BRS) machinery developed for Yang-Mills theories~\cite{BRS-75,BRS-76}. This leads to the introduction of a \emph{composite ghost} which handles the infinitesimal residual gauge freedom of the initial symmetry if any, and which enters into a ``Russian formula'' for the dressed connection.

\smallskip
The paper is organized as follows. In section~\ref{Dressing field method in a nutshell}, the geometrical framework underlying the dressing field method is briefly recalled. 

Then, in section~\ref{Composing the dressing fields}, 
the dressing of a conformal Cartan connection is performed in two steps with suitable dressing fields. For the normal conformal Cartan connection, we end with its so-called ``Riemannian parametrization'', which straightforwardly provides known geometric objects relevant in conformal geometry~ \cite{Ogiue}. 
Composition of dressing is shown to conspire in order to erase both the special conformal transformations (inversions) and the Lorentz $SO(1, m-1)$-gauge symmetry. The remaining symmetry turns out to be given by the simple abelian group $\mathsf{W}\simeq\doubleR_+\setminus \{0\}$ of Weyl rescalings. This step by step reduction rests on the fact that the two needed dressing fields satisfy some compatibility conditions and can ultimately be merged together.

Section~\ref{The dressing field and the BRS framework} investigates at the infinitesimal level the compatibility between the dressing field method and the elegant and efficient language of the BRS differential algebra~\cite{BRS-76,Sto76,Sto84}. A detailed account of the various modified BRS algebras coming out from the dressing scheme is given. Two examples illustrate this point. The first one concerns GR.
The second one is the BRS treatment (linearized version) of the finite procedure presented in the first part of the paper concerning the second order conformal structure. 

Section~\ref{conclusion} gathers some concluding remarks.

\section{The dressing field method in a nutshell}
\label{Dressing field method in a nutshell}

In this section, the main idea of the dressing scheme 
is introduced in a geometrical setting. To start with the basics, it is recalled that, the usual geometrical framework for dealing with Yang-Mills theories is that of a principal bundle $\P=\P(\M,H)$ over a space-time $\M$, with structure group~$H$. Let us denote by $\LieH$ its Lie algebra. Let $\omega \in \Omega^1(\P,\LieH)$ be a connection $1$-form on $\P$ and $\Omega \in \Omega^2(\P, \LieH)$ its curvature; 
let $\Psi$ denote a section of an associated bundle constructed out of a representation $(V, \rho)$ of~$H$.

Through a local trivializing section $\sigma:\U\subset \M \rarrow \P$, 
one gets the usual Yang-Mills gauge potential $A:=\sigma^*\omega$, the field strength $F:=\sigma^*\Omega$ and the matter field $\psi:=\sigma^*\Psi$. The local formulation on space-time will be understood with respect to the open set $\U$ throughout the paper.

\smallskip
To characterize the geometry of gauge fields, one has to specify the action of the (local) gauge group on each space of fields. The latter is defined by 
\[
\H = \big\lbrace
\gamma :\U\subset \M \rarrow H \big\rbrace ,
\]
with the group law inherited from $H$. The space $\H$ can also be considered as a space of gauge fields for the following action of $\gamma^{}_2 \in \H$ (as a group) on $\gamma^{}_1\in \H$ (as a space of fields):
\begin{equation}
\label{gamma_de_gamma}
{\gamma^{}_1}^{\gamma^{}_2} := \gamma_2\- \gamma^{}_1 \gamma^{}_2,
\end{equation}
which is compatible with the group law in $H$.

The field space  $\A$ of Yang-Mills potentials carries the usual action of the gauge group $\H$: $A\mapsto A^\gamma := \gamma\- A \gamma + \gamma\- d\gamma$, and accordingly, $F \mapsto F^\gamma:= \gamma\- F \gamma$. One has $\psi\mapsto \psi^\gamma := \rho(\gamma\-) \psi$ for matter fields.

In the following, we consider new fields composed of ``elementary'' fields. For instance, taking $A\in \A$ and $\gamma_1\in \H$ (as a space of fields), one constructs the ``composed field'' $A^{\gamma^{}_1}:= \gamma_1\- A \gamma^{}_1 + \gamma_1\- d\gamma^{}_1$. 
An induced action of the gauge group $\H$ is naturally defined by taking the gauge group action on each of the elementary fields.
As an illustration, for $A^{\gamma^{}_1}$ and for $\gamma_2 \in \H$ (as a group) one has
\begin{align}
\label{doublegaugetrsf}
{(A_{}^{\gamma^{}_1})}^{\gamma^{}_2} :\!&= {(A^{\gamma^{}_2})}^{{\gamma^{}_1}^{\gamma^{}_2}} 
= {(A^{\gamma^{}_2})}^{\gamma_2\- \gamma^{}_1 \gamma^{}_2} \notag\\[2mm]
&= (\gamma_2\- \gamma^{}_1 \gamma^{}_2)\- (A^{\gamma^{}_2}) (\gamma_2\- \gamma^{}_1 \gamma^{}_2) + (\gamma_2\- \gamma^{}_1 \gamma^{}_2)\- d(\gamma_2\- \gamma^{}_1 \gamma^{}_2) 
\notag\\[2mm]
& = (\gamma^{}_1 \gamma^{}_2)\- A (\gamma^{}_1\gamma^{}_2) + (\gamma^{}_1 \gamma^{}_2)\- d(\gamma^{}_1 \gamma^{}_2)
= A^{(\gamma^{}_1 \gamma^{}_2)}  \ ,
\end{align}
where  $A^{(\gamma^{}_1 \gamma^{}_2)}$ identifies with the action of $\gamma^{}_1 \gamma^{}_2\in\H$ (as a group) on $A\in\A$. 
Notice that this case is quite degenerate since $A^{\gamma^{}_1}\in \A$ and accordingly ${(A_{}^{\gamma^{}_1})}^{\gamma^{}_2} = {\gamma^{}_2}\- (A_{}^{\gamma^{}_1}) \gamma^{}_2 + {\gamma^{}_2}\- d\gamma^{}_2=A^{(\gamma^{}_1 \gamma^{}_2)}$. 
More subtle situations will be encoutered in the following.

\smallskip
The dressing field method involves identifying, in a local trivialization of $\P$, a local dressing field in $\D = \{u:  \U \rarrow G' \}$ where $G'$ is a target Lie group. This target group $G'$ has to be chosen compatible with $H$ and $\H$ and their representations. In particular, one requires that there exists a subgroup $H'\subseteq H$ such that $\D$ supports the following action of the subgroup $\H':=\{\gamma': \U \rarrow  H'\} \subseteq \H$:
\[
u \mapsto u^{\gamma'} := {\gamma'}^{-1} u.
\]
Contrary to~\eqref{gamma_de_gamma} this action is not compatible with the natural group law in $G'$.
The clear distinction between $\H$ and $\D$ as $\H'$-gauge field spaces, 
is crucial, and should be kept in mind.

Another requirement on $G'$ is the possibility to define, starting from the gauge fields $A, F, \psi,$ and $u$, the
following \emph{composite fields}: 
\begin{align}
\label{dressing}
\h A &:=A^u= u\- A u +  u\-du, & \h F &:=F^u= u\-F u, & & \text{and} &  \h\psi &:=\psi^u=\rho(u\-)\psi.
\end{align}
It can be checked that $\h F=d\h A + \tfrac{1}{2} [\h A,\h A]$. These composite fields are readily seen to be $\H'$-gauge invariant (\cite[Main Lemma]{GaugeInv}).

At this stage, two possibilities are in order. First, if $H' = H$ the above composite fields are $\H$-gauge invariant; the whole symmetry $\H$ has thus been erased.
Second, if $H'\subsetneqq H$, then the composite field might display a
residual gauge freedom: only the symmetry subgroup $\H'$ has been
neutralized. In either case, one can check indeed that, for instance (similarly to~\eqref{doublegaugetrsf})
 ${(A^u)}^{\gamma'} := {(A^{\gamma'})}^{u^{\gamma'}} = {(A^{\gamma'})}^{{\gamma'}\- u} = A^u$,  for any $\gamma' \in \H'\subseteq \H$.

\smallskip
This construction turns out to be a geometrical root of the notion of ``Dirac variables'' \cite{Dirac55, Dirac58} and its generalization to non-abelian gauge field theories. As shown in~\cite{GaugeInv} the method applies to the electroweak sector of the Standard Model (see also~\cite{Masson-Wallet}) and to the Einstein-Cartan formulation of GR. In both cases it provided an interpretive shift with respect to the usual viewpoint. 
In~\cite{NucleonSpin} we showed how the method ought to be at the root of the so-called ``Chen \& al. trick'' which has been sparking much work and reviving controversies on the nucleon spin decomposition issue. 

\section{Composing dressing fields}  
\label{Composing the dressing fields}

\subsection{The second order conformal structure}
\label{The second order conformal structure}

We refer to~\cite{Sharpe} for a detailed description of the Möbius geometry and its associated Cartan geometry.
See also~\cite{Kobayashi, Ogiue} for a formulation in terms of higher order frame bundles. Notice that it is of prior importance to have a link between the second order conformal structure and a matrix representation. In the latter, one gains a direct contact with the usual formulation of a Yang-Mills gauge theory. Also, the matrix formulation allows easier calculations. We just provide the basic material necessary to our construction. 

Let $\M$ be a $m$-dimensional smooth manifold ($m\geq 3$). Let $CO(\M)$ be the principal bundle of orthonomal frames with respect to the Minkowski $\eta$ metric of signature $(1,m-1)$\footnote{One could have worked with an arbitrary signature $(p,q)$, see {\em e.g.}~\cite{Friedrich:1995vb}.} with structure group
\[
K_0 \simeq CO(1,m-1) = \big\lbrace
M \in GL_m(\mathbb{R});\ M^T \eta M = z^2 \eta,\
z \in\mathsf{W} \big\rbrace = SO(1,m-1) \times \mathsf{W}
\]
where $\mathsf{W} = \mathbb{R}_+\!\!\setminus\! \{0\}$
 is the group of Weyl rescaling. The Lie algebra of $K_0$ is 
\[
\LieK_0:=\co(1,m-1)= \big\lbrace v \in~\gl_m(\mathbb{R});\ v^T\eta + \eta v = \epsilon \1_m,\
\epsilon \in\mathbb{R} \big\rbrace = \mathfrak{o}(1,m-1) \oplus \mathbb{R}.
\]
The principal bundle $CO(\M)$ is a reduction of $GL(\M)$ the principal bundle of linear frames over $\M$. This is a first order $G$-structure which can be prolongated to a second order $G$-structure~\cite{Kobayashi}. The latter is easier recast in the the following setting.

The Klein model geometry is the pair of Lie groups $(G, H)$ where $G= O(2,m) / \lbrace \pm I_{m+2} \rbrace$ with
\[
 O(2,m) = \left\{ M \in GL_{m+2}(\doubleR) \ |\  M^T\Sigma M =\Sigma, \mbox{ where } 
\Sigma= \begin{pmatrix} 0 & 0 & -1 \\ 0 & \eta & 0 \\  -1 & 0 & 0 \end{pmatrix} \right\} . 
\]
It is the isometry group of the de Sitter space $dS^m$ defined as the quadric $\Sigma(x,x)=0$ in $\mathbb{R}^{m+2}$.
$H$ is the isotropy group of the point $(1,0,\dots,0)$ such that $dS^m \simeq G/H$ and has the following factorized matrix representation
\begin{align*}
 H = K_0\, K_1=\left\{ \begin{pmatrix} z &  0 & 0  \\  0  & S & 0 \\ 0 & 0 & z^{-1}  \end{pmatrix}  \begin{pmatrix} 1 & r & \frac{1}{2}rr^t \\ 0 & \1 & r^t \\  0 & 0 & 1\end{pmatrix}\  \bigg|\ z\in \mathsf{W},\ S\in SO(1, m-1), 
\ r\in \doubleR^{m*} \right\},
\end{align*} 
where ${}^t$ stands for the $\eta$-transposition, namely for the row vector $r$ one has $r^t = (r \eta^{-1})^T$ (where ${}^T\,$ is the usual matrix transposition) and $\doubleR^{m*}$ is the dual of $\doubleR^m$. Notice that $K_1$ is an abelian subgroup of $H$. It can be shown that $H \simeq CO(1,m-1)  \ltimes \doubleR^{m*}$.

The infinitesimal Klein pair is $(\LieG, \LieH)$, where both are graded Lie algebras~\cite{Kobayashi}. They can respectively be decomposed  according to $\LieG=\LieG_{-1}\oplus\LieG_0\oplus\LieG_1 \simeq \doubleR^m\oplus\co(1, m-1)\oplus\doubleR^{m*}$, a splitting which gives the different symmetry sectors: translations + (Lorentz $\times$ Weyl) + inversions, and $\LieH=\LieG_0\oplus\LieG_1 \simeq \co(1,m-1)\oplus\doubleR^{m*}$. The quotient space is just $\LieG/\LieH=: \LieG_{-1}\simeq \doubleR^m$. In matrix notation we have,
\begin{multline*}
\mathfrak{g}=\left\{ 
\begin{pmatrix} \epsilon &  \iota & 0  \\  \tau  & v & \iota^t \\ 0 & \tau^t & -\epsilon  \end{pmatrix} \bigg|\ (v-\epsilon\1)\in \mathfrak{co}(1, m-1),\ \tau\in\mathbb{R}^m,\ \iota\in\mathbb{R}^{m*}  
\right\} \\
\supset
\LieH = \LieG_0\oplus\LieG_1 = \left\{ \begin{pmatrix} \epsilon &  \iota & 0  \\  0  & v & \iota^t \\ 0 & 0 & -\epsilon  \end{pmatrix}\bigg| \textrm{\ldots} \right\},
\end{multline*} 
with the $\eta$-transposition $\tau^t = (\eta\tau)^T$ of the  column vector $\tau$.
The graded structure of the Lie algebras, $[\LieG_i, \LieG_j] \subseteq \LieG_{i+j}$, $i,j=0,\pm 1$ with the abelian Lie subalgebras $[\LieG_{-1}, \LieG_{-1}] = 0 = [\LieG_1, \LieG_1]$, is automatically handled by the matrix commutator.

The second order conformal structure modelled on this Klein pair is a principal bundle, $\P(\M, H)$,  with structure group $H$, together with a (local) Cartan connection $\varpi \in \Omega^1(\U , \LieG)$ with curvature $\Omega=d\varpi+\varpi^2\  \in \Omega^2(\U, \LieG)$. Accordingly, in matrix representation, the Cartan connection is parametrized by the matrix of $1$-forms
\[
\varpi= \begin{pmatrix}  a & \alpha & 0 \\ \theta & A &\ \alpha^t \\  0 &\ \theta^t & -a  \end{pmatrix}\ ,
\]
where $\theta$ is the soldering (or vielbein) 1-form which gives an isomorphism between 
each tangent space $T_x\M$ and $\LieG_{-1} \simeq \doubleR^m$;  while the curvature is given by
\[
\Omega= \begin{pmatrix} f & \Pi & 0 \\ \Theta & F & \Pi^t \\ 0 & \Theta^t & -f  \end{pmatrix} := \begin{pmatrix} da + \alpha\theta & d\alpha +\alpha(A -a\1) & 0   \\   d\theta +(A-a\1)\theta &\ \ dA + A^2 + \theta\alpha + \alpha^t\theta^t &\ \ d\alpha^t + (A+a\1)\alpha^t \\  0 &  d\theta^t + \theta^t(A +a\1) &  -da +\theta^t\alpha^t \end{pmatrix},
\]
where the wedge product of forms is tacitly assumed.

\smallskip
The \emph{normal conformal Cartan connection} is the unique $\varpi$ (up to $\H$-gauge transformations) whose curvature is constrained by the two conditions~\cite{Ogiue,Kobayashi,Sharpe}
\begin{align}
\label{NC3}
& i)\ \Theta \equiv 0, \mbox{ (torsion-free geometry)}, &
& ii)\ \textrm{Ric}(F)= {F^a}_{bac}\equiv 0, \mbox{ (Ricci-null condition)},
\end{align}
where the Einstein summation convention is understood and will be used throughout the paper. Combination of the latter with $\LieG_{-1}$-sector of the Bianchi identity $d \Omega + [\varpi,\Omega]=0$ yields the traceless condition $f=0$.

\smallskip
An element $\gamma$ of the gauge group $\H$ can be factorized as
\begin{align*}
\gamma=\gamma_0\gamma_1: \U \rarrow H=K_0\, K_1, \quad\textrm{with } \left\{
    \begin{array}{l}
   \gamma_0\in \K_0 := \big\lbrace \gamma :\U\rarrow K_0 \big\rbrace ,\\[2mm]
    \gamma_1 \in \K_1:= \big\lbrace  \gamma :\U\rarrow K_1 \big\rbrace .
    \end{array}
  \right.
\end{align*}
Accordingly, with respect to $\K_0$, the gauge transformations of the Cartan connection are, 
\begin{align}
\label{GT_0}
\varpi^{\gamma_0}& =\gamma_0\-\varpi\gamma_0 + \gamma_0\-d\gamma_0 \notag \\[2mm]
& =: \begin{pmatrix}
a^{\gamma_0} & \alpha^{\gamma_0} & 0 \\ {\theta^{}}^{\gamma_0} & {A^{}}^{\gamma_0} & (\alpha^{\gamma_0})^t \\  0 & ({\theta^{}}^{\gamma_0})^t & - a^{\gamma_0} 
\end{pmatrix}
= \begin{pmatrix} a+ z\-dz & z\-\alpha S &  0 \\ S\-\theta z & S\-AS +S\-dS & S\-\alpha^t z\-  \\  0  & z\theta^t S &  -a +zdz\-  \end{pmatrix},
\end{align}
and with respect to $\K_1$,
\begin{multline}
\label{GT_1}
\varpi^{\gamma_1}= \gamma_1\-\varpi\gamma_1 + \gamma_1\-d\gamma_1 
=: \begin{pmatrix}
a^{\gamma_1} & \alpha^{\gamma_1} & 0 \\ {\theta^{}}^{\gamma_1} & {A^{}}^{\gamma_1} & (\alpha^{\gamma_1})^t \\  0 & ({\theta^{}}^{\gamma_1})^t & - a^{\gamma_1} 
\end{pmatrix}
\\[2mm]
= \begin{pmatrix} a-r\theta &\ ar - r\theta r +\alpha -rA +\frac{1}{2}rr^t\theta^t +dr &  0 \\ \theta  & \theta r + A - r^t\theta^t &\ \theta \frac{1}{2}rr^t + Ar^t -r^t\theta^t r^t + \alpha^t + r^t a + dr^t\\  0  & \theta^t  &  \theta^t r^t-a  \end{pmatrix}.
\end{multline}

The principal bundle $\P(\M, H)$ is a second order $G$-structure, a reduction of the second order frame bundle $L^2\M$; it is thus a ``2-stage bundle''. The bundle $\P(\M,H)$ over $\M$ can also be seen as a principal bundle $\P_1:=\P(\P_0,K_1)$ with structure group $K_1$ over $\P_0:=\P(\M,K_0)$, see~\cite{Kobayashi}.
The whole structure group $H= K_0\, K_1\simeq CO(1,m-1) \ltimes \doubleR^{m*}$ is of dimension $1+\frac{m(m-1)}{2}+m.$

As shown in the next section, the symmetry group $H$ can be reduced to the $1$-dimensional Weyl group $\mathsf{W}$ through the dressing field method.

\subsection{The need for two dressing fields}
\label{2dressings}

The second order bundle geometry $\P_1 \rarrow \P_0 \rarrow \M$ suggests that the structure group $K_1$ should be neutralized first, in order to reach the principal bundle $\P_{\mathsf{W}}:=\P(\M, \mathsf{W})$.

\paragraph{First reduction.}  
In order to neutralize $K_1$ (inversions) and so to `reduce' $H=K_0\, K_1$ to the $K_0$ factor, we seek a dressing field, that is a local map, see~\cite{JF_PhD}, 
\begin{align*}
u_1: \U \rarrow K_1 \quad\textrm{ such that }\quad u_1^{\gamma^{}_1}=\gamma_1\-u_1,
\mbox{ for any } \gamma^{}_1 \in \K_1,
\end{align*}
whose matrix expression is
\[
 u_1=\begin{pmatrix} 1 &\ q & \frac{1}{2}qq^t \\ 0 &\ \1 & q^t \\  0 & 0 & 1\end{pmatrix},
\]
where $q: \U \rarrow \doubleR^{m*}$ is a covector field. 

Such a dressing field can be extracted from a  `gauge-fixing-like' constraint, $\chi(\varpi^{u_1})=0$, imposed on the dressed Cartan connection $\varpi_1$. The vanishing of the $(1,1)$-component (thus $(3,3)$ as well) of $\varpi^{u_1}$ is taken to be such a constraint. 
On account of \eqref{GT_1}, it explicitly reads
\[
\chi(\varpi^{u_1}):\   a^{u_1} = a - q\theta = 0.
\]
Solving the constraint for $q$ leads to
\[
a-q\theta=a-q_a\theta^a= a_\mu dx^\mu -q_a {e^a}_\mu dx^\mu=0\ \Rightarrow
q_a=a_\mu {(e\-)^\mu}_a, 
\]
or in index free notation $q=a\!\cdot\! e\-$, where ``$\cdot$'' means greek index summation and will be used throughout the paper.
In the latter, some care should be taken, bearing in mind that $a$ is a covector, the \textit{scalar coefficients} of the $1$-form $a$. The distinction should be clear according to the context. Remark that the contraint yields a solution $q$ which is \textit{local}  in the field theory sense, (\textit{i.e.} a differential polynomial in the fields), contrary to the non local realisation of a dressing field as can be found in~\cite{Dirac55,Dirac58}.

Now, the $\K_1$-gauge transformation of the Cartan connection \eqref{GT_1} gives us,
\begin{align*}
a^{\gamma_1}&=a-r\theta \quad \rarrow (a^{\gamma_1})_\mu=a_\mu -r_a {e^a}_\mu, \quad \textrm{or in index free notation} \quad a^{\gamma_1}=a -re,\\
\theta^{\gamma_1}&=\theta \quad \rarrow \quad {(e^{\gamma_1})^a}_\mu={e^a}_\mu \quad \textrm{in index free notation} \quad e^{\gamma_1}=e.
\end{align*}
This implies $q^{\gamma_1}= a^{\gamma_1}\!\cdot\!{e\-}^{\gamma_1}=(a-re)\cdot e\-=a\!\cdot\! e\- -r=q-r$. The other two entries of $u_1^{\gamma_1}$ are computed to be $(q^t)^{\gamma_1}=q^t - r^t$, and  $\frac{1}{2}(qq^t)^{\gamma_1}=\frac{1}{2}(qq^t + rr^t) -rq^t$. 

The $\K_1$-transformations of $u_1$ is {\em dictated} by the $\K_1$-gauge transformations of the gauge potentials \emph{i.e.} the entries of the Cartan connection matrix. It is thus remarkable that they precisely allow $u_1$ to be truly a dressing field, as can be seen in matrix notation,
\begin{align}
\label{u_1_1}
u_1^{\gamma_1}=\gamma_1\-u_1= \begin{pmatrix} 1 &\ -r &\ \frac{1}{2}rr^t \\ 0 &\ \1 & - r^t \\  0 &\ 0 & 1\end{pmatrix}\, \begin{pmatrix} 1 &\ q &\ \frac{1}{2}qq^t \\ 0 &\ \1 & q^t \\  0 &\ 0 & 1\end{pmatrix}=\begin{pmatrix} 1 &\ q-r &\ \frac{1}{2}(qq^t+rr^t) -rq^t \\ 0 & \1 & q^t-r^t \\  0 & 0 & 1\end{pmatrix}.
\end{align}
It is worthwhile to notice, on the contrary, that for $\gamma_1, \gamma'_1 \in \K_1$, a simple matrix calculation shows that $\gamma'_1{\,}^{\gamma_1}:= \gamma_1\- \gamma'_1\, \gamma_1=\gamma'_1$, since $K_1$ is abelian.

Having now at our disposal the dressing field $u_1$, we can proceed to the dressing of the Cartan connection, in accordance with~\eqref{dressing}. One readily computes
\begin{align}
\label{varpi_1}
\varpi_1:\!&=\varpi^{u_1}  = u_1\-\varpi u_1 + u_1\-d\u_1  \notag\\[2mm]
\begin{pmatrix} 0 & \alpha_1^{} & 0  \\ \theta & A_1^{} & \alpha_1^t \\ 0 &  \theta^t & 0  \end{pmatrix} &= \setlength{\arraycolsep}{7pt}
\begin{pmatrix} a-q\theta &\ (a -q\theta) q +\alpha -qA +\frac{1}{2}qq^t\theta^t +dq &  0 \\
\theta  & \theta q + A - q^t\theta^t & \big(\mbox{entry }(1,2)\big)^t 
\\  0  & \theta^t  &  \theta^t q^t-a  \end{pmatrix}\ , 
\end{align}
where, by construction, $a_1=a-q\theta \equiv 0$. Likewise, for the dressed curvature,
\begin{align}
\label{Omega_1}
\Omega_1:\!&=\Omega^{u_1} = u_1\- \Omega u_1 \notag\\[2mm]
\begin{pmatrix}  f_1 &\ \Pi_1 & 0 \\ \Theta_1 &\ F_1 &\ \Pi_1^t \\ 0 &\ \Theta_1^t & -f_1 \end{pmatrix}& =\, \setlength{\arraycolsep}{6pt}
\begin{pmatrix} f-q\Theta &\ \Pi - qF_1 + fq -\frac{1}{2}qq^t\Theta^t & 0  \\  \Theta & 
\Theta q + F -q^t\Theta^t &  \big(\mbox{entry }(1,2)\big)^t 
\\  0 & \Theta^t & \Theta^t q^t -f \end{pmatrix}. 
\end{align}
By construction, $\varpi_1$ and $\Omega_1$ are $\K_1$-gauge invariant composite fields. This means in particular that the expression of $\varpi_1$ is invariant, that is $a^{}_1=0=a_1^{\gamma_1}$.
It is worthwhile to notice that in~\cite{Sharpe} the condition $a\equiv 0$ is considered as a gauge fixing, named ``natural gauge'' in~\cite{Korzynski:2003bh}, while it actually emerges through a dressing.

\medskip
Let us now study the behaviour of the composite field $\varpi_1$ under Lorentz transformations,
$SO(1, m-1) \subset K_0$, since any element $\gamma_0\in \K_0$ is factorized as
\begin{align*}
\gamma_0=W S := \begin{pmatrix} z & 0 &  0 \\  0 & \1 & 0 \\ 0 & 0 &
  z\- \end{pmatrix} \, \begin{pmatrix} 1 & 0 & 0 \\ 0 &S & 0 \\ 0 & 0
  & 1 \end{pmatrix}, 
\end{align*}
where $z\in \mathsf{W}$ and $S\in SO(1,m-1)$ --with $S^T \eta S=\eta$-- has been identified with its matrix-block representation.
The action of the Weyl subgroup $\mathsf{W}$ is treated seperately in Appendix~\ref{Appendix}.

By its very construction, $\varpi_1^S$ depends on the Lorentz gauge transformations of the Cartan connection $\varpi$ (set $z=1$ in~\eqref{GT_0}) 
\begin{equation*}
\varpi^S = S\-\varpi S + S\-dS = \begin{pmatrix} a & \alpha S &  0 \\ S\-\theta  &\ S\-AS +S\-dS &\ S\-\alpha^t   \\  0  & \theta^t S &  -a \end{pmatrix}.
\end{equation*}
Since $\theta^S=S\-\theta=e^S \!\cdot\! dx=S\- e \!\cdot\! dx$, we get $q^S= a^S\!\cdot\! (e\-)^S= a\!\cdot\!(e\-)S=qS$, $(q^S)^t = (qS)^t=S\-q^t$ and thus $(qq^t)^S = qq^t$. Hence, the transformed $u_1^S$ of $u_1$ under the Lorentz group relies on those of the Cartan connection entries. Once more, it is remarkable that in matrix notation it can be recast as
\begin{align}
\label{u_1_0}
u_1^S=S\- u_1 S =\begin{pmatrix} 1 & 0 & 0 \\ 0 & S\- & 0 \\ 0 & 0 &1 \end{pmatrix}\, \begin{pmatrix} 1 & q & \frac{1}{2}qq^t \\ 0 & \1 & q^t \\  0 & 0 & 1\end{pmatrix} \, \begin{pmatrix} 1 & 0 & 0 \\ 0 & S & 0 \\ 0 & 0 &1 \end{pmatrix}  = \begin{pmatrix} 1 & qS & \frac{1}{2}qq^t \\ 0 & \1 & S\-q^t \\  0 & 0 & 1\end{pmatrix}.
\end{align}
The latter shows that the $K_1$-valued dressing field $u_1$ is subject   
to a Lorentz gauge-like transformation.
Therefore, according to a calculation similar to~\eqref{doublegaugetrsf}, the composite fields $\varpi_1$ and $\Omega_1$ are Lorentz gauge transformed as
\begin{align}
\label{Omega_1^S}
\varpi_1^S = S\- \varpi_1 S +S\-dS, \qquad\quad \Omega_1^S=S\-\Omega_1 S.
\end{align}
This allows to conclude that the entries of $\varpi_1$ and $\Omega_1$ are \textit{true} $\mathcal{SO}$-gauge fields. This means that $A_1$ is the true Lorentz/spin connection with curvature $R_1:= dA_1+A_1^2$. 
To sum up, $\varpi_1$ and $\Omega_1$ are $K_1$-invariant composite fields but still remain $\mathcal{SO}$-gauge fields. 

\smallskip
Furthermore, if $\varpi$ is chosen to be the \emph{normal} conformal Cartan connection~\eqref{NC3}, then $\varpi_1$ satisfies similar normality conditions:
\begin{align}
\label{NCC_1}
\Theta_1 = 0,\qquad \textrm{Ric}(F_1)=0.
\end{align}
Indeed from~\eqref{Omega_1} one has $\Theta_1=\Theta=0$ and
$\textrm{Ric}(F_1)=\textrm{Ric}(\Theta q + F - q^t\Theta^t)=\textrm{Ric}(F)=0$. Furthermore, the trace-free condition $f_1=f-q\Theta=0$ is obvious.
We thus get
\begin{align*}
\varpi_1=\begin{pmatrix} 0 & \alpha_1 & 0 \\ \theta & A_1 & \alpha_1^t \\ 0 & \theta^t & 0\end{pmatrix}, \quad \textrm{and} \quad \Omega_1=\begin{pmatrix} 0 & \Pi_1{} & 0 \\ 0 & F_1^{} & \Pi_1^t \\ 0 & 0 & 0  \end{pmatrix}=\begin{pmatrix} 0 & d\alpha_1^{} + \alpha_1^{}A_1^{}& 0 \\ 0 &\ R_1+ \theta\alpha_1^{}+\alpha^t_1\theta^t &\ d\alpha_1^t  +A_1^{} \alpha^t_1 \\ 0 & 0 & 0\end{pmatrix},
\end{align*}
where $A_1$ is still the \emph{Lorentz/spin-connection}, $\alpha_1$ may be referred to as the \emph{Schouten} $1$-form (by solving in $\alpha_1$ the equation $ \textrm{Ric}(F_1)=0$), $R_1$ is the \emph{ Riemann curvature} $2$-form, $F_1$ may be called the \emph{Weyl curvature} $2$-form, and finally, $\Pi_1$ may be named the \emph{Cotton} $2$-form, that is the covariant differential of the Schouten $1$-form with respect to the spin-connection $A_1$. 

\medskip
The ${\cal SO}$-gauge fields $\varpi_1$ and $\Omega_1$ (normal or not) are associated with the \textit{first-order structure} $\P_0:=\P(\M, K_0)=CO(\M)$, since $SO\subset K_0$. It then makes sense to ask whether the Lorentz gauge symmetry  can be neutralized by finding an adequate \textit{second} dressing field. This is indeed possible as shown in the following. 

\paragraph{Second reduction.}   
We want to neutralize the Lorentz subgroup of $K_0$, leaving only the abelian Weyl subgroup as the final residual gauge symmetry. 
The suitable dressing field is extracted from $\varpi^S$, or $\varpi_1^S$. Indeed, we have $\theta^S=S\-\theta$ which provides the transformation law for the vielbein $e\in GL_m(\doubleR)$,  $e^S=S\-e$. Hence, one may define the local map 
\begin{align}
& u_0^{}:\U \rarrow GL_{m+2}(\doubleR)\supset K_0,\quad  \textrm{with matrix form, }\quad  u_0=\begin{pmatrix} 1 & 0 & 0 \\ 0 & e & 0 \\ 0 &  0 & 1 \end{pmatrix},\quad \textrm{such that} \notag\\
 & u_0^S=S\-u_0^{} \quad \rarrow \quad \begin{pmatrix} 1 & 0 & 0 \\ 0 & e^S & 0 \\ 0 &  0 & 1 \end{pmatrix}=\begin{pmatrix} 1 & 0 & 0 \\ 0 & S\- & 0 \\ 0 &  0 & 1 \end{pmatrix}\, \begin{pmatrix} 1 & 0 & 0 \\ 0 & e & 0 \\ 0 &  0 & 1 \end{pmatrix}=\begin{pmatrix} 1 & 0 & 0 \\ 0 & S\-e & 0 \\ 0 &  0 & 1 \end{pmatrix}. \label{u_0_0}
\end{align}
Note that such a field $u_0$ is valued in a larger group than the one to be erased, $G'\!\simeq GL_m \supset SO=H'$.
Moreover, from $\varpi^{\gamma^{}_1}$, see~\eqref{GT_1}, one extracts
\begin{align}
\label{u_0_1}
\theta^{\gamma_1}=\theta \quad \rarrow \quad  e^{\gamma_1}=e, \quad \textrm{which implies} \quad u_0^{\gamma_1}=u_0.
\end{align}
Eqs \eqref{u_0_0} and \eqref{u_0_1} secure that $\varpi_1$ can be dressed by $u_0$. This gives rise to a new composite field
which is inert under both the $\K_1$- and ${\cal SO}$-actions:
\begin{align}
\label{final varpi}
\varpi_0 & := \varpi_1^{u_0}=u_0\-\varpi_1 u_0^{} + u_0\-du_0^{} \notag\\[2mm]
& =: \begin{pmatrix} 0 & P & 0 \\  dx & \Gamma & g\-\!\!\cdot\! P^T \\ 0 & dx^T \!\!\cdot\! g & 0 \end{pmatrix}
= 
\begin{pmatrix} 0  & \alpha_1 e  & 0  \\  e\-\theta &\ e\-A_1 e + e\-de &\ e\-\alpha_1^t \\ 0 & \theta^t e & 0 \end{pmatrix},
\end{align}
where $g$ is the metric on $\M$ induced by the Cartan connection through $e^T\eta e=g$. In the last matrix equality, $\Gamma:=e\-A_1 e +e\-de$ and $P:=\alpha_1^{}e$ are \textit{definitions}, the other entries are directly obtained from the calculation. In great detail
\begin{align*}
e\-\theta &= e\- e \!\cdot\! dx = \delta \!\cdot\! dx = dx, \\
\theta^t e &= \theta^T \eta e = dx^T \!\cdot\! e^T\ \eta e =dx^T \!\!\cdot\! g, \\
e\-\alpha_1^t &= e\- \eta\-\alpha_1^T = g\- \!\!\cdot\! e^T\alpha_1^T=g\- \!\!\cdot\!(\alpha_1^{}e)^T=g\- \!\!\cdot\! P^T. 
 \end{align*} 
In components, $\varpi_0$ thus reads
 \begin{align}
 \label{final varpi 2}
 \varpi_0 = \begin{pmatrix} 0 & P_{\mu\nu} & 0 \\  \delta_\mu^\rho &\ {\Gamma^\rho}_{\mu\nu}&\ g^{\rho\lambda}P_{\lambda\mu} \\ 0 & g_{\mu\nu} & 0 \end{pmatrix} dx^\mu,
\end{align}
Actually, the transformation under coordinate changes of $\varpi_0$, in part due to $u_0$, allows to identify $\Gamma$ as a linear connection $1$-form. It turns out that $\varpi_0$ is parametrized by geometric objects on $\M$.
The curvature associated to $\varpi_0$ is the following $\K_1$- and ${\cal SO}$-invariant composite field
 \begin{align}
\label{Omega_0}
 \Omega_0:= \Omega_1^{u_0}=u_0\-\Omega_1 u_0 =\begin{pmatrix} f_1 & \Pi_1 e & 0 \\ e\-\Theta &\ e\- F_1e &\ e\-\Pi^t \\ 0 & \Theta^t e  &  -f_1\end{pmatrix} =: \begin{pmatrix} f_0 & C & 0 \\ T & W & C^t \\ 0 & T^t & -f_0  \end{pmatrix}.
 \end{align}
Instead, by computing $\Omega_0= d\varpi_0 + {\varpi_0}^2$ directly, the above matrix explicitly reads
\begin{align}
\label{final Omega}
\begin{pmatrix} f_0 & C & 0 \\ T & W & C^t \\ 0 & T^t & -f_0  \end{pmatrix} 
= 
\begin{pmatrix}  P\wedge dx &  dP + P\wedge \Gamma & 0  \\  \Gamma\wedge dx  &\ \ R +dx\wedge P + g\-\!\!\cdot\! P^T \wedge dx^T\!\!\cdot\! g &\ \ \nabla g\- \wedge P^T +g\-\!\!\cdot\! C^T \\ 0 & -dx^T\wedge \big( \nabla g +\Gamma^T\!\!\cdot\! g\big) & dx^T\wedge P^T  \end{pmatrix}, 
\end{align}
where $R=d\Gamma +\Gamma^2$ is the curvature of the linear connection $\Gamma$ on $\U$,  and $\nabla g = dg -\Gamma^Tg-g\Gamma$ is the covariant derivative of the metric with respect to the linear connection.  The metricity condition $\nabla g=0$ is automatically satisfied as can be checked due to the fact that $A_1$ is $\so$-valued.
Expressing~\eqref{final Omega} in components reads
\begin{align}
\Omega_0 & = \tfrac{1}{2}\begin{pmatrix} (f_0)_{\mu\sigma} & C_{\nu, \mu\sigma} & 0 \\  {T^\rho}_{\mu\sigma} &\ {W^\rho}_{\nu, \mu\sigma} & {C^\rho}_{\mu\sigma} \\ 0 & T_{\nu, \mu\sigma} & -(f_0)_{\mu\sigma}  \end{pmatrix} dx^\mu\wedge dx^\sigma  \notag \\[2mm] 
\label{final Omega 2}
& = \begin{pmatrix}  P_{[\mu,\sigma]} &  \partial_{[\mu} P_{\sigma],\nu} + P_{[\mu,\lambda}{\Gamma^\lambda}_{\sigma],\nu} & 0 \\[1.5mm]  {\Gamma^\rho}_{[\mu,\sigma]} &  \   \tfrac{1}{2}{R^\rho}_{\nu, \mu\sigma} + \delta^\rho_{[\mu} P_{\sigma],\nu}^{} + P_{[\mu,\lambda}\, g^{\lambda\rho}  g_{\sigma]\nu}  &\  \nabla_{[\mu} g^{\rho\lambda}P_{\lambda,\sigma]} +g^{\rho\lambda} \tfrac{1}{2} C_{\lambda, \mu\sigma} \\[1.5mm]  0 &  \nabla_{[\mu} g_{\sigma]\nu} +{\Gamma_{[\mu,\sigma]}}^\lambda g_{\lambda\nu}  &  -P_{[\mu,\sigma]}  \end{pmatrix} dx^\mu\wedge dx^\sigma .
\end{align}

\bigskip
Restricting ourselves to the {\em normal} case~\eqref{NC3}, $\varpi_0$ provides the so-called \emph{Riemannian parametrization of the normal conformal Cartan connection}~\cite{Ogiue}.
The normality conditions on $\varpi_0$ defined by:
\begin{align*}
T = 0,  \qquad \text{Ric}(W) =0.
\end{align*}
are fulfilled.
Indeed, from both~\eqref{Omega_0} and~\eqref{final Omega 2} one has
\begin{align}
\label{NC3_Riemann}
T=e\-\Theta_1=0,\quad  \mbox{ and }\quad
\text{Ric}(W) :=\text{Ric}(e\-F_1 e) = \text{Ric}(e\- F e)=\text{Ric}(F)=0,
\end{align}
and hence $f_0=f_1=P\wedge dx=0$.

Accordingly, $T=0$ implies the symmetry of $\Gamma$ in its lower indices, and it can be shown in the usual way that $\Gamma$ can be expressed as a function of $g$ in order to get the \emph{Levi-Civita connection} on $\U$. In addition, the condition $f_0=0$ renders $P_{\mu\sigma}$ symmetric which is nothing but the so-called \emph{Schouten tensor}, so that $C_{\nu, \mu\sigma}=\nabla_\mu P_{\sigma\nu}$ is the \emph{Cotton tensor} and ${W^\rho}_{\nu, \mu\sigma}$ is the \emph{Weyl tensor}. 
To sum up, in the \emph{normal} case we have,
\begin{align}
\label{NCC}
 \varpi_0 &= \begin{pmatrix} 0 & P_{\mu\nu} & 0 \\  \delta_\mu^\rho &\ {\Gamma^\rho}_{\mu\nu}& g^{\rho\lambda}P_{\lambda\mu} \\ 0 & g_{\mu\nu} & 0 \end{pmatrix} dx^\mu,\quad \textrm{with} \quad  P_{\mu\nu} = \frac{-1}{(m-2)}\bigg( R_{\mu\nu} - \frac{R}{2(m-1)}g_{\mu\nu} \bigg), \\
\label{NCC_curv}
 \Omega_0&=\tfrac{1}{2}\begin{pmatrix} 0 & C_{\nu, \mu\sigma} & 0 \\  0 & \ {W^\rho}_{\nu, \mu\sigma} & g^{\rho\lambda}C_{\lambda, \mu\sigma} \\ 0 & 0  & 0   \end{pmatrix} dx^\mu\wedge dx^\sigma.
\end{align}
Formulae \eqref{NCC} and \eqref{NCC_curv} are known as the Riemannian parametrization of the normal conformal Cartan connection.
In full generality, without assuming the normality conditions, the composite field given in~\eqref{final varpi} thus provides a more general parametrization of the conformal Cartan connection. 

\medskip
The composite fields $\varpi_0$ and $\Omega_0$ (normal or not) 
are now associated to the Weyl bundle $\P_{\mathsf W}$.
Remark that this step of reducing $\P_0$, with structure group $K_0=\mathsf{W} \times SO \simeq CO(1, m-1)$, to $\P_{\mathsf W}^{}$ with abelian structure group $\mathsf{W}$, is quite analogous to the case of the electroweak sector of the Standard Model where the initial bundle with structure group $SU(2)\times U(1)$ was reduced to a subbundle with abelian structure group $U(1)$; see~\cite{Masson-Wallet,GaugeInv}.

\paragraph{Two steps in one: the compatibility conditions.}   
   
It is possible to go from $(\varpi, \Omega)$ to $(\varpi_0, \Omega_0)$ in a single step, because the dressing fields obey necessary compatibility conditions 
which are the last two equations in the following collected set
\begin{align*}
& u_1^{\gamma_1}=\gamma_1\-u_1, \quad u_0^S=S\-u_0 \quad \textrm{equation \eqref{u_1_1} and \eqref{u_0_0} respectively.}\\
& u_1^S=S\-u_1S, \quad u_0^{\gamma_1}=u_0 \quad \textrm{equation \eqref{u_1_0} and \eqref{u_0_1}.}
\end{align*}
These imply,
\begin{align*}
\big(u_1u_0\big)^{\gamma_1}&= u_1^{\gamma_1} u_0^{\gamma_1}=\gamma_1\-u_1  u_0 =\gamma_1\- \big(u_1u_0\big),\\
\big(u_1u_0\big)^S &= u_1^S\, u_0^S=S\-u_1S S\-u_0=S\-\big(u_1u_0\big),
\end{align*}
which show that the double dressing field $u_1u_0$ is a dressing for the groups $\K_1$ and ${\cal SO}$ respectively. 
The above two equations say more, since they imply (in full detail), 
\begin{align*}
\big(u_1u_0\big)^{S\gamma_1}&=\bigg( \big(u_1u_0\big)^S\bigg)^{\gamma_1}=\bigg(S\- \big(u_1u_0\big)\bigg)^{\gamma_1}=(S^{\gamma_1})\- \big( u_1u_0\big)^{\gamma_1}\\
& = \gamma_1\-S\-\gamma_1^{} \gamma_1\-\big(u_1u_0\big) =  \gamma_1\-S\- \big(u_1u_0\big)=(S\gamma_1)\-\big(u_1u_0\big),
\end{align*}
so that $u_1u_0$ turns out to be a dressing field for the whole gauge subgroup ${\cal SO}\, \K_1$. 

\smallskip
To sum up, had we guessed the sole composite dressing field $u \defeq u_1u_0: \U \rarrow K_1\, GL_{m+2}(\doubleR)$, we could have 
directly dressed the Cartan connection and its curvature in order to get at once the composite fields $\varpi_0:=\varpi^u$ and $\Omega_0:=\Omega^u$, constructed above. 

\smallskip
Since $\varpi_0$ and $\Omega_0$ are associated to the \emph{Weyl bundle} $\P_{\mathsf W}^{}$, a \emph{residual Weyl gauge symmetry} is expected to hold. The next section addresses this issue.

\subsection{The residual  Weyl  symmetry}   
\label{Residual  Weyl  symmetry}   

The  goal of our analysis is to obtain the transformations of the final composite fields $\varpi_0$ (eq.\eqref{final varpi}) and $\Omega_0$ (eq.\eqref{final Omega}) under the residual Weyl symmetry. These residual transformations arise from the \emph{combined} transformations under the Weyl group of the Cartan connection $\varpi$  and its curvature $\Omega$ on the one hand, and, of the two dressing fields, on the other hand. Indeed, one is led to define the Weyl transformation of the fields as
\begin{align}
\label{comp-field-W}
\varpi_0^W := \big( \varpi^{u_1u_0} \big)^W= \big( \varpi^W \big)^{u_1^W u_0^W}, \quad \text{and }\quad \Omega_0^W := \big( \Omega^{u_1u_0} \big)^W = \big(\Omega^W \big)^{u_1^W u_0^W}.
\end{align}
The Weyl transformation of the Cartan connection is,
\begin{align}
\label{CartanC-Weyl}
\varpi^W = W\-\varpi W + W\-dW = \begin{pmatrix}
a+ z\-dz &\ z\-\alpha & 0 \\
\ z\theta & A & \alpha^tz\- \\
0 & z\theta^t & \ -a +zdz\-
\end{pmatrix},
\end{align}
from which we can see that $\theta^W = z\theta\ \rarrow\ e^W=ze$. Turning this into a matrix notation, one has
\begin{align}
\label{u_0_W}
u_0^W :=  \begin{pmatrix} 1 & 0  & 0 \\ 0 & e^W  &  0 \\ 0 & 0 & 1\end{pmatrix}
=
\tilde W u_0 := \begin{pmatrix} 1 & 0  & 0 \\ 0 & z  &  0 \\ 0 & 0 & 1\end{pmatrix} \, \begin{pmatrix} 1 & 0  & 0 \\ 0 & e   &  0 \\ 0 & 0 & 1\end{pmatrix}=\begin{pmatrix} 1 & 0  & 0 \\ 0 & ze  &  0 \\ 0 & 0 & 1\end{pmatrix}\  ,
\end{align}
where $\tilde W$ is thus \textit{another matrix representation} of the Weyl group, adapted to the dressing field $u_0$, different than the initial representation $W$.

\smallskip
The Weyl tranformation of the dressing $u_1$ stems from \eqref{CartanC-Weyl}: $a^W= a +z\-dz$, where $a$ is a $1$-form, so that in covector notation $a^W=a +z\- \partial z =: a + \zeta$, where $a$ now stands for the scalar coefficients of the $1$-form $a:=a \!\cdot\! dx$.
Then, given $u_1 \sim q:=a \!\cdot\! e\-$, one computes
\begin{align}
\label{u_1_W}
q^W & := a^W\cdot(e\-)^W= (a+\zeta) \!\cdot\! z\-e\- =z\-\big( q + \zeta \!\cdot\! e\- \big),
\end{align}
likewise, $(q^t)^W = (q^W)^t$. One readily verifies $(qq^t)^W = q^W (q^W)^t$.
These transformation laws suggest that the abelian composition law in
the subgroup $K_1$ must enter into the game. Having this point in mind, a rather
tricky matrix expression for the transformed field $u_1^W$ is found out
\begin{align}
\label{u_1^W}
\MoveEqLeft
u_1^W =
\begin{pmatrix}
1 & q^W & \frac{1}{2}q^W (q^W)^t \\[1mm] 0 & \1 & (q^W)^t \\  0 & 0 & 1 
\end{pmatrix} = 
\begin{pmatrix}
1 &\ z\-\big( q + \zeta \!\cdot\! e\- \big) &\ \tfrac{1}{2} z^{-2} \big( q + \zeta\!\cdot\! e\- \big) \big( q + \zeta\!\cdot\! e\- \big)^t \\[1mm]
0 & \1 & z\- \big( q + \zeta\!\cdot\! e\- \big)^t \\
0 & 0 & 1
\end{pmatrix} \\
& = W\- u_1 k_1 W = W\- k_1 u_1 W 
:= W\-\, u_1\, \begin{pmatrix} 
1 & \ \zeta\!\cdot\! e\- &\ \frac{1}{2} \zeta \!\cdot\! e\-  (\zeta \!\cdot\! e\-)^t \\
0 & \1 & (\zeta \!\cdot\! e\-)^t \\  0 & 0 & 1
\end{pmatrix}\, W \notag
\end{align}
which suggests to proceed as after~\eqref{u_1_0} in order to get the Weyl transformations of the composite fields.

 \paragraph{Residual Weyl gauge symmetry of $\bs{\varpi_0}$.}  
 
In virtue of the definition~\eqref{comp-field-W}, for the composite field $\varpi_0$ we have, 
\begin{align*}
\varpi_0^W :\!&=
\big( \varpi^W \big)^{u_1^W u_0^W}= \big( \varpi^W \big)^{W\-u_1 k_1 W{\tilde W}u_0} \\
&= \varpi^{u_1k_1 u_0 W{\tilde W}} = \varpi^{u_1 u_0 u_0\-k_1 u_0 W{\tilde W}} = {\varpi_0}^{u_0\-k_1 u_0 W{\tilde W}} \\
& =: {\b W}\- {\varpi_0}{\b W}  + {\b W}\- d {\b W}.
\end{align*}
where we have used the fact that $[W{\tilde W},u_0] = 0$ and defined the $z$-dependent matrix
\[
\b W := u_0\-k_1 u_0 W{\tilde W} =
\begin{pmatrix}
z &\quad z \zeta &\quad  \tfrac{1}{2} z\- \zeta \!\cdot\! g\-  \!\cdot\! \zeta^T \\[2mm]
0 &\quad z \delta\ &\quad z\- g\-  \!\cdot\! \zeta^T \\
0 &\quad 0 &\quad z\-
\end{pmatrix}. 
\] 
A straightforward calculation yields
\begin{multline}
\label{Weyl final varpi}
 \begin{pmatrix} 0 & P^W & 0 \\ dx^W & \Gamma^W &  (g\- P^T)^W \\ 0 & (dx^T\!\!\cdot\! g)^W & 0 \end{pmatrix} =\\ 
\setlength{\arraycolsep}{3pt}
\begin{pmatrix} 
0 & P + \nabla\zeta - \zeta \!\cdot\! dx \zeta + \tfrac{1}{2} \zeta \!\cdot\! g\- \!\!\cdot\! \zeta^T dx^T \!\!\cdot\! g & 0 \\ 
dx & \Gamma + z\- dz \delta + \zeta dx - g\- \!\!\cdot\! \zeta^T dx^T \!\cdot\! g & z^{-2} g\- \!\!\cdot\! \big(\mbox{entry }(1,2)\big)^T \\ 
0 &\quad dx^T\!\!\cdot\! z^2 g & 0 
\end{pmatrix}.
\end{multline}
Let us elaborate on this result.
Entry (2, 1) gives $dx^W=dx$, which is the obvious invariance of the coordinate chart under a Weyl rescaling. Entry (3, 2) gives,
\begin{align}
\label{metric-W}
(dx^T\cdot g)^W =  dx^T \cdot g^W = dx^T\cdot (z^2g),\quad \text{in components} \quad \big( g_{\mu\nu}\big)^W =z^2g_{\mu\nu},
\end{align}
which is the \emph{Weyl rescaling of the metric tensor} under the conformal factor $z^2$. 
 Entry (2, 2) gives in components
\begin{align}
\label{Christoffel-W2}
 \big({\Gamma^\rho}_{\mu\nu} \big)^W = {\Gamma^\rho}_{\mu\nu} + \delta^\rho_\nu\  \zeta_\mu   + \delta^\rho_\mu\ \zeta_\nu  -   g^{\rho\lambda}\ \zeta_\lambda \  g_{\mu\nu},
\end{align}
with $\zeta_\mu:= z\-\d_\mu z$. 
Entry (1, 2) gives in components,
\begin{align}
\label{Schouten-W}
\big(P_{\mu\nu}\big)^W  = P_{\mu\nu} + \nabla_\mu\zeta_\nu  -\zeta_\mu \ \zeta_\nu   + \tfrac{1}{2}\,  \zeta_\lambda \zeta^\lambda \ g_{\mu\nu} .
\end{align}
with $\zeta^\lambda:= g^{\lambda\alpha}\zeta_\alpha $ . 
Entry (2, 3) gives in components,
\begin{align*}
 \big(g^{\rho\lambda} P_{\lambda\mu}\big)^W= z^{-2} g^{\rho\lambda}\big(  P_{\lambda\mu}  +  \nabla_\mu\zeta_\lambda   - \zeta_\lambda \zeta_\mu  +  \tfrac{1}{2}\,  g_{\lambda\mu}\  \zeta_\alpha \zeta^\alpha      \big) ,
\end{align*}
which is redundant with \eqref{metric-W} and \eqref{Schouten-W} respectively. 

\medskip
\noindent\textbf{\Rmk}:$\quad$ Equations \eqref{Christoffel-W2} and
\eqref{Schouten-W} look like the familiar conformal transformations of
the Christoffel symbols and of the Schouten tensor. Notice however
that in this framework, the metricity condition $\nabla g=0$ being
guaranteed, $\Gamma$ reduce to the Levi-Civita connection in the
{\em normal} case ($T=0$) {\em only}. We will return to that specific
case down below.
However, the above calculations hold even without restricting
ourselves to this assumption. We then obtain at once the Weyl
variation of both the symmetric \textit{and} anti-symmetric parts of $\Gamma$ and $P$. Explicitly, decomposing ${\Gamma^\rho}_{\mu\nu} = {\Gamma^\rho}_{[\mu\nu]}+{\Gamma^\rho}_{(\mu\nu)}$ and $P_{\mu\nu}= P_{[\mu\nu]} +P_{(\mu\nu)}$, then for the Christoffel symbols of the linear connection
\begin{equation}
\label{AS-W}
\big( {\Gamma^\rho}_{[\mu\nu]} \big)^W = {\Gamma^\rho}_{[\mu\nu]}, \qquad\textrm{and}\qquad \big({\Gamma^\rho}_{(\mu\nu)} \big)^W= {\Gamma^\rho}_{(\mu\nu)}\  +\ \delta^\rho_\nu\  \zeta_\mu \  +\ \delta^\rho_\mu\ \zeta_\nu \  -\   g^{\rho\lambda}\ \zeta_\lambda \  g_{\mu\nu},
\end{equation}
while for the coefficients of the $1$-form $P$ one gets
\begin{align}
\label{P-W}
\big( P_{[\mu\nu]} \big)^W &=  P_{[\mu\nu]} - \zeta_\lambda {\Gamma^\lambda}_{[\mu\nu]} \notag\\[-2.5mm]
& \\[-2.5mm]
 \text{and}\qquad  \big(P_{(\mu\nu)}\big)^W &= P_{(\mu\nu)} +
 \d_{(\mu} \zeta_{\nu)} - \zeta_\lambda {\Gamma^\lambda}_{(\mu\nu)}
 - \zeta_\mu \, \zeta_\nu + \tfrac{1}{2}\, \zeta_\lambda \zeta^\lambda \, g_{\mu\nu}, \notag
\end{align}
where $ \d_{(\mu} \zeta_{\nu)} = \d_\mu \zeta_\nu$ from the very definition of $\zeta$.
The two equalities concerning the
symmetric parts are nothing but the \emph{transformations of the
  Christoffel symbols and of the Schouten tensor under Weyl rescaling
  of the metric}. 

\smallskip
Thus, with~\eqref{Christoffel-W2}
and~\eqref{Schouten-W}, not only do we recover classical results in a
much more effective way, but we have more: 
we do not need to assume a priori that $\Gamma$ and $P$ are functions of
the metric~$g$, as it is usually the case when one works with the
Levi-Civita connection.
It is pretty noticeable to find out these general Weyl transformations as a `top-down' process.

\paragraph{Residual Weyl gauge symmetry of $\bs{\Omega_0}$.}
 
Similarly, for the composite field $\Omega_0$ (eq.\eqref{final Omega}), formula \eqref{comp-field-W} gives
\[
\Omega_0^W  := \big(\Omega^W \big)^{u_1^W u_0^W} = \big(\Omega^W \big)^{W\-u_1 k_1
  W{\tilde W}u_0} = \Omega^{u_1 u_0 u_0\-k_1 u_0 W{\tilde W}}=: {\b W}\- {\Omega_0}{\b W},
\]
\begin{multline}
\label{Weyl final Omega}
\begin{pmatrix}
f_0^W & C^W & 0 \\
T^W &  W^W & {C^t}^W \\ 
0 & {T^t}^W & -f_0^W 
\end{pmatrix} = \\[2mm]
 \begin{pmatrix}
f_0 - \zeta \!\cdot\! T &\ \ C - \zeta \!\cdot\! W + (f_1 - \zeta \!\cdot\!
T)\zeta + \tfrac{1}{2} \zeta\!\cdot\! g\- \!\cdot\! \zeta^T T^T
\!\cdot\! g & 0\\
T & W + T\zeta - g\- \!\cdot\! \zeta^T T^T \!\!\cdot\! g &\ z^{-2} g\-
\big(\mbox{entry }(1,2)\big)^T
\\
0 & T^T \!\cdot\! z^2g & (\zeta \!\cdot\! T)^T - f_0
\end{pmatrix}. 
\end{multline}
 In components, entries (1, 1) and (3, 3) give
\begin{align}
\label{trace-W}
\big(2 P_{[\mu,\sigma]}\big)^W= 2P_{[\mu,\sigma]} - \zeta_\lambda {T^\lambda}_{\mu\sigma}, 
\end{align}
which just reproduces the first equation in \eqref{P-W} above, since $ \tfrac{1}{2}{T^\lambda}_{\mu\sigma}= {\Gamma^\lambda}_{[\mu,\sigma]}$.\\
Entry (2, 1) gives,
\begin{align}
\label{torsion-W}
\big({T^\rho}_{\mu\sigma}\big)^W = {T^\rho}_{\mu\sigma} 
\end{align}
which reproduces the first equation in \eqref{AS-W}. \\
Entry (3, 2) is redundant with \eqref{torsion-W} and \eqref{metric-W}. 
Entry (2, 2) gives,
\begin{align}
\label{pseudoWeyl-W}
\big({W^\rho}_{\nu, \mu\sigma}\big)^W = {W^\rho}_{\nu, \mu\sigma} + {T^\rho}_{\mu\sigma}\, \zeta_\nu - g^{\rho\lambda} \zeta_\lambda  \, {T_{\mu\sigma}}^\alpha g_{\alpha\nu}
\end{align}
Entry (1, 2) leads to
 \begin{align}
 \label{pseudoCotton-W}
 \big(C_{\nu, \mu\sigma}\big)^W = C_{\nu, \mu\sigma} - \zeta_\lambda {W^\lambda}_{\nu, \mu\sigma} + \zeta_\lambda\ \delta^\lambda_\nu\  f_{\mu\sigma} +  \tfrac{1}{2} ( \zeta_\lambda g^{\lambda\alpha} \zeta_\alpha  ) {T_{\mu\sigma}}^\beta g_{\beta\nu} -  \zeta_\lambda   {T^\lambda}_{\mu\sigma}\, \zeta_\nu.
\end{align}
At last, entry (2, 3) is redundant with \eqref{metric-W} and \eqref{pseudoCotton-W}. 

\paragraph{Residual Weyl gauge freedom of $\bs{\varpi_0}$ and $\bs{\Omega_0}$ in the normal case.}   

It is now straightforward to specialize the above transformations to the case where the initial Cartan connection $\varpi$ is normal. 
The dressed normal conformal Cartan connection $\varpi_0$ has been given in~\eqref{NCC}, with curvature~\eqref{NCC_curv}.
It is readily seen from the Weyl variation $\Omega_0^W$ given in~\eqref{Weyl final Omega} that the normality of $\varpi_0$, as stated by~\eqref{NC3_Riemann}, is preserved under the action of the Weyl gauge group.

Formula~\eqref{Weyl final varpi} remains formally unchanged but $\Gamma$ becomes the Levi-Civita connection and $P$ is the Schouten tensor, while the Weyl variation of the curvature~\eqref{Weyl final Omega} reduces to
\begin{align*}
\Omega_0^W = 
\begin{pmatrix}
0 &\ C^W & 0 \\
0 &\  W^W &\ {C^W}^t \\ 
0 & 0 & 0 
\end{pmatrix}  
= \begin{pmatrix}
0 &\ C - \zeta \!\cdot\! W & 0\\
0 & W &\ z^{-2} g\- \big( C - \zeta \!\cdot\! W \big)^T \\
0 & 0 & 0
\end{pmatrix}.
\end{align*}
The Weyl rescaling of the metric tensor is still given by~\eqref{metric-W}.
But~\eqref{Christoffel-W2} and~\eqref{Schouten-W} now expresses respectively 
the transformations of the Levi-Civita connection and of the Schouten tensor under Weyl rescaling. 
Equation~\eqref{pseudoWeyl-W} reduces to
\begin{align}
 \big({W^\rho}_{\nu, \mu\sigma}\big)^W = {W^\rho}_{\nu, \mu\sigma}
\end{align}
and is the well known \emph{invariance of the Weyl tensor under Weyl rescaling}. Finally, \eqref{pseudoCotton-W} gives
\begin{align}
 \big(C_{\nu, \mu\sigma}\big)^W = C_{\nu, \mu\sigma} - \zeta_\lambda {W^\lambda}_{\nu, \mu\sigma},
\end{align}
which is the \emph{transformation of the Cotton-York tensor under Weyl rescaling}.

\begin{table}[t]
\centering
\resizebox{\textwidth}{!}{
\begin{tabular}{l | cc| cc}
\toprule
& \multicolumn{2}{c|}{Starting geometry} & \multicolumn{2}{c}{Outcoming geometry}\\
 & Second order 
& Degrees of freedom & Natural geometry &  Degrees of freedom \\
& conformal structure
&  &  &   \\
\hline \midrule
& & & & \\
(1) Variables & $\varpi \in \Omega^1(\U ,\LieG)$ & $m (1+\frac{m(m-1)}{2} + 2m )$ & $g_{\mu\nu}$ & $ \frac{m(m+1)}{2}$ \\[3mm]
 &   & &  ${\Gamma^\rho}_{\mu\nu}$  &  $m^3$ \\[1mm]
 & & &$P_{\mu\nu}$ & $0$ \\
& & & & \\
\midrule
& & & & \\
(2)  Symmetries &  $ 
(SO \times \mathsf{W})\, \doubleR^{m*}$ & $\frac{m(m-1)}{2} + 1+ m$ & $\mathsf{W}=\mathbb{R}_+\!\!\setminus\! \{0\}$ & $1$ \\
& & & & \\
\midrule
& & & & \\
(3) Contraints  &  $\Theta=0$ & $m\, \frac{m(m-1)}{2}$ & $\nabla g=0$ & $m\,\frac{m(m+1)}{2}$ \\[3mm]
 & \text{Ric}(F)=0 &	$\frac{m(m+1)}{2}$ & $T=0$ &  $ m\, \frac{m(m-1)}{2}$ \\[3mm]
 &$f=0$  & $\frac{m(m-1)}{2}$ & & \\
& & & & \\
\midrule
& & & & \\
$\begin{array}{l}
\mbox{Total degrees} \\
\mbox{of freedom}\\
(1)-(2)-(3)
\end{array}$ & & $\frac{m(m+1)}{2} -1$ & & $\frac{m(m+1)}{2} -1$ \\
& & & & \\
\bottomrule
\end{tabular}
}
\caption{Counting the degrees of freedom of the \emph{normal} conformal geometry before and after the dressing operation. 
Finally, remark that the normal geometry is the most natural one for its total degrees of freedom are those of a conformal class $[g]$ of the metric $g$. }
\label{normal-case}
\end{table}

\newpage

\section{The dressing field and the BRS framework}  
\label{The dressing field and the BRS framework}  

Since its inception in the mid 70’s by Becchi, Rouet and Stora in~\cite{BRS-75,BRS-76}, the
BRS formalism has seeded considerable work, and has been generalized to a large class of theories  and became a standard tool in the analysis of gauge field theories and  their quantization.

In the following, we give the minimal definition of the BRS algebra of a theory and show how the latter is modified through the dressing field method. Then we apply the construction to General Relativity (GR) and to the second order conformal structure.

\subsection{Residual BRS algebra}  
\label{Residual BRS algebra}  

Consider the bundle $\P(\M, H)$ with a (local) connection $A$ together
with  a section $\psi$ of any associated bundle. Given the
$\LieH$-valued ghost $v$, the BRS algebra is, 
\begin{align*}
sA = -Dv, \qquad sF=\big[F, v\big], \qquad s\psi=-\rho_*(v)\psi, \qquad \text{and}  \qquad sv=-\tfrac{1}{2}[v, v],
 \end{align*}  
with $D=d\ +[\omega,\ \ ]$, and $F = d A + \tfrac{1}{2} [A,A] $. The commutator $[\alpha, \beta] =\alpha\beta - (-)^{|\alpha|\,|\beta|}\beta\alpha$ is graded according to total degree which consists of the form and ghost degrees. $\rho_*$ is the representation of the Lie algebra $\LieH$ on matter fields. The BRS operator, $s$, is nilpotent: $s^2=0$. 

\noindent 
This can be compactly rewritten under the elegant algebraic equation, the so-called ``Russian formula''~\cite{Sto84} (also named ``horizontality condition''~\cite{Ne'eman:1978gg,Baulieu:1981sb})
\begin{equation}
\label{eq-russe}
 (d+s)(A+v) + \tfrac{1}{2} [A+v,A+v] = F\ ,
\end{equation}
according to an expansion with respect to the ghost degree. The sum $A+v$ is named the ``algebraic connection''~\cite{DbV86}.

\smallskip \noindent
As a first basic result, one states:
 \begin{lem}[Modified BRS algebra]\label{Modified BRS algebra}    
Let $u: \U \rarrow\ G'$ be a field as in section~\ref{Dressing field method in a nutshell} on which the action of the gauge group $\H$ is not specified yet. Let 
 \begin{align*}
 \h A:=  u\- A u+ u\-du, \quad \h F=u\- F  u, \quad \textrm{and}\quad \h\psi:=\rho( u\-)\psi,
 \end{align*}
be the corresponding composite fields. Then there is a modified  BRS algebra: 
\begin{align}
\label{modified-BRS-algebra}
s\h A = -\h D\h v, \qquad s\h F=\big[\h F, \h v\big], \qquad s\h\psi = -\rho_*(\h v)\h\psi, \qquad \text{and}  \qquad s\h v=-\tfrac{1}{2}[\h v,\h v],
\end{align}
with new ghost given by
\begin{align}
\label{composite-ghost}
\h v =  u\- v  u +   u\- s  u .
\end{align}
\end{lem}

\smallskip \noindent
This \emph{composite ghost}  is the central new variable. To the best of our knowledge, first appearances of such an object in specific examples can be found in~\cite{Garajeu-Grimm-Lazzarini,Lazzarini-Tidei}.

\begin{proof} ${}$\\   
It is easily checked by expressing each variable of the initial BRS algebra as function of the composite variables and the dressing field. In the course of the checking, the explicit expression for the variation $su$ is not required.
\end{proof}

\smallskip \noindent
Accordingly, we have the following

\begin{cor} \label{Modified Russian formula}  
One can define a composite algebraic connection,
\begin{align}
\h A + \h v= u\-\big(A+v\big) u + u\-(d+s)u
\end{align}
which, by virtue of the above modified BRS algebra, satisfied the modified Russian formula, 
\begin{align}
(d+s)\big( \h A + \h v\big) + \tfrac{1}{2}\big[ \h A +\h v, \  \h A +\h v\big] =\h F.
\end{align}
\end{cor}

\paragraph{Three relevant possibilities.}  

The modified BRS algebra of Lemma \ref{Modified BRS algebra} can take various presentations according to explicit expression of the \emph{composite ghost} which is dictated by a given BRS transformation of the field $ u:\U \rarrow G'$. In this respect, three cases will be considered.

\medskip
First case. Suppose $u$ is subject to a \emph{gauge-like} finite $\H$-transformation, that is $ u^\gamma=\gamma\-  u\gamma$, for $\gamma \in \H$. Then its BRS transformation, mimicking infinitesimal gauge transformations with the ghost $v$ as $\LieH$-valued parameter, is 
\begin{align} 
s u=\big[  u, v\big].
\end{align}
This implies that the ghost is kept unchanged $\h v = v$ and
 \eqref{modified-BRS-algebra} reads
\begin{align}
\label{cas1}
s\h A = -\h D v, \qquad s\h F=\big[\h F, v\big], \qquad s\h\psi = -\rho_*(v)\h\psi, \qquad \text{and}  \qquad 
s v=-\tfrac{1}{2}[ v, v].
\end{align}
This is not a surprise since if $u$ were a gauge element $ u \in \H$, the fields $\h A$, $\h F$ and  $\h \psi$ would actually be the $\H$-gauge transformed of $A$, $F$ and $\psi$ respectively, satisfying the same BRS algebra.

\medskip
Second case. 
Suppose that $u$ is a \emph{dressing field for $\H$} that is $u^\gamma=\gamma\- u$, for $\gamma \in \H$. Then its BRS transformation is,
\begin{align}
s u= - v\, u.
\end{align}
This implies that the \textit{composite ghost} vanishes $\h v = 0$.
The $\H$-symmetry is thus annihilated and the \textit{modified BRS algebra} \eqref{modified-BRS-algebra} reduces to the \emph{trivial algebra},
\begin{align}
\label{trivial-BRS}
s\h A=0,\qquad s\h F=0, \qquad \text{and} \qquad s\h\psi=0.
\end{align}
This expresses that $\h A$, $\h F$ and $\h\psi$ are \emph{$\H$-gauge invariants} fields. In that case, it is stressed that the composite fields coming out from the dressing field method ought to be good candidates for being ``observables''.

\medskip
Third case. Let $H' \subset H$ be a Lie subgroup with Lie algebra $\LieH'$.
Suppose that $u$ transforms according to $ u^{\gamma'}= {\gamma'}\- u$ for $\gamma' \in \H'$. For the time being, we left unspecified the transformation $ u^{\gamma_0}$ for  $\gamma_0 \in \H \setminus \H'$. Suppose we can find a  $\Ad_H$-invariant complement, $\Liep$, to $\LieH'$ in $\LieH$, so that $\LieH=\LieH' \oplus \Liep$. 
The ghost splits according to $v=v_{\LieH}=v_{\LieH'} +v_{\Liep}$ and accordingly the BRS operator splits too as $s=s_{\LieH}=s_{\LieH'}+s_{\Liep}$  with
\begin{align}
\label{splitBRS}
& s^2=0\ \Leftrightarrow\ s_{\LieH'}^2 =  s_{\Liep}^2 = s_{\LieH'}\, s_{\Liep} + s_{\Liep}\, s_{\LieH'} = 0\ , \notag \\[-2.5mm]
& \\[-2.5mm]
sv= - \tfrac{1}{2} [v,v] \ \Leftrightarrow\ & s_{\LieH'} v_{\LieH'} =  - \tfrac{1}{2} [v_{\LieH'},v_{\LieH'}], \quad
s_{\LieH'} v_{\Liep} =  - [v_{\LieH'},v_{\Liep}],\quad
s_{\Liep} v_{\Liep} =  - \tfrac{1}{2} [v_{\Liep},v_{\Liep}]. \notag
\end{align}
The BRS transformation of the dressing field is then,
\begin{align}
\label{su_residuel}
s u=s_{\LieH} u= s_{\LieH'}  u + s_{\Liep} u= -v_{\LieH'}  u+ s_\Liep  u,
\end{align}
where $s_\Liep u$ is left unspecified. This implies for the \textit{composite ghost},
\begin{align}
\label{final-ghost}
\h v &= u\- v_\LieH  u +  u\-s_\LieH u = u\-v_{\LieH'} u + u\-v_\Liep u - u\-v_{\LieH'} u +  u\-s_\Liep  u \notag\\
&= u\-v_\Liep u + u\-s_\Liep  u=:\h v_\Liep,
\end{align}
showing that the whole $\H'$ subsector has been neutralized. 
The composite ghost $\h v_\Liep$ encodes the residual gauge symmetry. Thus, the \textit{modified BRS algebra} \eqref{modified-BRS-algebra} reads 
\begin{align}
\label{reduced-BRS}
s\h A =-\h D \h v_\Liep, \qquad s\h F= \big[ \h F, \h v_\Liep  \big], \qquad s\h\psi=-\rho_*(\h v_\Liep)\h\psi, \qquad \text{and} \qquad s\h v_\Liep =-\tfrac{1}{2}\big[ \h v_\Liep, \h v_\Liep  \big],
\end{align}
and gives the \emph{infinitesimal residual $\H/\H'$ gauge transformations} of the composite fields $\h A$, $\h F$ and $\h\psi$. This expresses the \emph{reduction} of the BRS algebra we started with.

\smallskip
These three cases cover pertinent types of erasure (none, total, partial) 
of gauge symmetry.
Let us apply these mechanisms to GR, on the one hand, and to the second order conformal structure, on the other hand.

\subsection{Application to the geometry of General Relativity}  
\label{Application to the geometry of General Relativity}  

The geometry underlying GR considered as a gauge theory, is a Cartan geometry $(\P, \varpi)$, where $\P(\M, H)=SO(\M)$ is a principal bundle with $H=SO(1,m-1)$ the Lorentz group and $\varpi \in \Omega^1(\U, \LieG)$ is a (local) Cartan connection on $\U\subset \M$ with values in $\LieG$ the Lie algebra of the Poincaré group $G=SO \ltimes \doubleR^m$.  One has the matrix writing
\begin{align*}
\varpi=\begin{pmatrix} A &\ \theta \\ 0 & 0 \end{pmatrix}=\begin{pmatrix} {A^a}_{b, \mu} & {e^a}_\mu \\ 0 & 0 \end{pmatrix}dx^\mu,
\end{align*}
with $A \in \Omega^1(\U, \LieH)$ the Lorentz connection and $\theta\in \Omega^1(\U, \doubleR^m)$ the vielbein $1$-form. The greek indices are spacetime indices, while latin indices are ``internal'' (gauge)-Minkowski indices. 
The curvature is 
\begin{align*}
\Omega=d\varpi+\tfrac{1}{2}[\varpi, \varpi]=d\varpi +\varpi^2 \quad \rarrow \quad \begin{pmatrix} F &\ \Theta \\ 0 & 0 \end{pmatrix}= \begin{pmatrix} dA+A^2 & \ d\theta+A\,\theta \\ 0 & 0 \end{pmatrix},
\end{align*}
with $F$ the Riemann $2$-form and $\Theta$ the torsion $2$-form. In this matrix notation, the Lorentz ghost reads
$v=\begin{pmatrix} v_L & 0 \\ 0 & 0 \end{pmatrix}$ and the Lorentz BRS algebra is
\begin{align*}
s\varpi &= - Dv, & s\Omega & = [\Omega, v], &  sv = -\tfrac{1}{2}& [v, v] = -v^2, \\[2mm]
\begin{pmatrix} sA & s\theta \\ 0 & 0 \end{pmatrix}&= \begin{pmatrix} -Dv_L &\ -v_L\theta \\ 0 & 0 \end{pmatrix}, & 
\begin{pmatrix} sF & s\Theta \\ 0 & 0 \end{pmatrix}&= \begin{pmatrix} [F, v_L] &\ -v_L\Theta \\ 0 & 0 \end{pmatrix}, &
 \begin{pmatrix} sv_L & 0 \\ 0 & 0 \end{pmatrix} &= \begin{pmatrix} -v_L^2 & \theta \\ 0 & 0 \end{pmatrix}.
\end{align*}
Of course $s^2=0$. This matrix algebra handles the infinitesimal $\SO$-gauge transformations of the variables of the theory. 

\medskip
As proposed in~\cite{GaugeInv} the dressing field is the vielbein, $u = \begin{pmatrix} e &  0 \\ 0 & 1 \end{pmatrix}:\U \rarrow GL_m(\doubleR)$. The corresponding composite fields express as
\begin{align*}
\h\varpi=u\-\varpi u+u\-du = \begin{pmatrix} e\-Ae +e\-de &\ e\-\theta \\ 0 & 0 \end{pmatrix}=: \begin{pmatrix} \Gamma & dx \\ 0 & 0  \end{pmatrix},
\end{align*}
where $\Gamma$ is a linear connection compatible with the metric defined by $g=e^T \!\eta e$, 
and
\begin{align*}
\h\Omega=u\-\Omega u= \h D\h\varpi=d\h\varpi +\h\varpi^2=\begin{pmatrix} d\Gamma +\Gamma^2 &\ \Gamma \!\cdot\! dx\\ 0 & 0\end{pmatrix} =: \begin{pmatrix}  R &\ T \\ 0 & 0  \end{pmatrix},
\end{align*}
where $ R$ is the Riemann tensor and $T$ is the torsion tensor. 

\smallskip
The key element of the modified BRS algebra is of course the composite ghost. To find its expression we only need to determine how the field $u$  transforms under the action of the (initial) BRS operator. It is readily read from $s\varpi$ above: $s\theta =-v_L\theta\  \rarrow \ se\cdot dx= -v_Le\cdot dx$. Hence, 
\[
su=-vu \ \rarrow\ \begin{pmatrix} se & 0 \\ 0 & 0  \end{pmatrix}=\begin{pmatrix} -v_Le & 0 \\ 0 & 0 \end{pmatrix},
\]
which is of course the defining BRS transformation of a dressing field.
Therefore the composite ghost vanishes by
\begin{align*}
\h v=u\-vu + u\-su = u\-vu +u\-(-vu)=0,
\end{align*}
and we have the trivial modified BRS algebra,
\begin{align*}
s\h\varpi= \begin{pmatrix} s\Gamma & sdx \\ 0 & 0 \end{pmatrix}=0, \qquad s\h \Omega=\begin{pmatrix} s R & sT \\ 0 & 0 \end{pmatrix}=0.
\end{align*}
This expresses the invariance of the coordinate chart, 
and by construction the $\SO$-gauge invariance of $\Gamma$, $R$ and $T$. 
The composite fields $\h \omega$ and $\h  \Omega$ belong to the natural geometry of $\M$. They are blind to the initial Lorentz gauge symmetry since the latter has been fully neutralized 
by the dressing field $u$. GR illustrates case two discussed above.
Let us now turn to the following less trivial example.


\subsection{Application to the second order conformal structure}  
\label{Application to the conformal structure}  

According to section~\ref{The second order conformal structure}, the structure group of the second order conformal structure $\P(\M, H)$ is $H=K_0\, K_1= (SO \times \mathsf{W})\, K_1$. Turning the infinitesimal parameters $(v_L,\epsilon,\iota)$ into Faddeev-Popov ghosts for the Lie algebra $\co(1,m-1)\oplus\doubleR^{m*}= \so(1,m-1) \oplus \doubleR\oplus\doubleR^{m*}$, the matrix-wise ghost decomposes into symmetry sectors ($L$ for Lorentz, $\mathsf{W}$ for Weyl and $i$ for the inversions) as
\begin{equation}
\label{ghost_v}
v = \begin{pmatrix} \epsilon & \iota & 0 \\ 0 & v_L & \iota^t \\ 0 & 0 & -\epsilon
\end{pmatrix} = v_0+v_i = v_L +v_i + v_{\mathsf{W}}= \begin{pmatrix} 0 & \iota & 0 \\ 0 & v_L & \iota^t \\ 0 & 0 & 0
\end{pmatrix} + \begin{pmatrix} \epsilon & 0 & 0 \\ 0 & 0 & 0 \\ 0 & 0 & -\epsilon
\end{pmatrix},
\end{equation}
where the Lorentz ghost $v_L$ is identified with its matrix block representation.
The BRS operation splits accordingly as
\[
s=s_0+s_i = s_{\mathsf{W}} + s_L +s_i \ ,
\]
and fulfils~\eqref{splitBRS}.

With the $\K_1$-dressing field $u_1:\U \rarrow K_1$, see~\eqref{u_1_1}, in addition to the composite fields $\varpi_1=~\varpi^{u_1}$, $\Omega_1=\Omega^{u_1}$, one constructs the first composite ghost
\begin{align*}
\h v_1:=v^{u_1}&=u_1\- v u_1 + u_1\- s u_1 = u_1\- \big(v_{\mathsf{W}}+v_L+v_i\big) u_1 + u_1\-\big(s_{\mathsf{W}}+s_L+s_i  \big)u_1.
\end{align*}
Upon linearizing the finite transformations~\eqref{u_1_1},~\eqref{u_1_0} and~\eqref{u_1^W}, one obtains the following BRS variations of the dressing field $u_1$
 \begin{align}
 \label{CC_1}
\begin{array}{l}
 s_i u_1 = - v_i u_1,\\[3mm]
s_L u_1 =\big[u_1, v_L\big],
\end{array}
\qquad \mbox{and} \qquad
s_{\mathsf{W}} u_1 = \begin{pmatrix} 0 & -\epsilon q + \d \epsilon\cdot e\- & -\epsilon qq^t +\d\epsilon\cdot e\-q^t \\ 0 & 0 & (-\epsilon q + \d \epsilon\cdot e\-)^t \\ 0 & 0 &0 \end{pmatrix},
 \end{align}
which illustrate the three cases discussed in section~\ref{Residual BRS algebra}. The first {\em composite ghost} is then  \begin{align}
 \h v_1 &= u_1\-v_{\mathsf{W}}u_1 + u_1\-v_Lu_1 +u_1\- v_i u_1 + u_1\-s_{\mathsf{W}}u_1 + u_1\-\big[u_1, v_L\big] -u_1\- v_i u_1 \notag\\
&= u_1\- v_{\mathsf{W}} u_1+ u_1\- s_{\mathsf{W}} u_1 + v_L \notag\\[2mm]
\label{first-comp-ghost}
&=\begin{pmatrix} \epsilon &\  \d \epsilon\!\cdot\! e\- & 0 \\ 0 & v_L & (\d \epsilon\!\cdot\! e\-)^t \\ 0 & 0 &-\epsilon \end{pmatrix}\ .
\end{align}
In the course of the computation, the ghost $v_i$ for the inversion sector has been killed by the dressing $u_1$ so that the subalgebra corresponding to $s_i$ is now trivial,
 \begin{align}
 s_i \omega_1 = 0 \quad\text{and} \quad  s_i\Omega_1=0, 
 \end{align} 
and expresses the expected $\K_1$-invariance of the composite fields $\varpi_1$ and $\Omega_1$. Thanks to the dressing $u_1$, the BRS operator $s=s_0+s_i=s_{\mathsf{W}}+s_L+s_i $ reduces to $s_0=s_{\mathsf{W}}+s_L$, and the Lorentz subalgebra is kept unchanged, 
 \begin{align*}
 s_L\varpi_1=-D_1v_L \quad \text{and} \quad s_L\Omega_1=\big[ \Omega_1, v_L \big],
 \end{align*}
 where $D_1 = d + \big[\varpi_1, \ \ \big]$ is the covariant derivative with respect to $\varpi_1$.
 This means that $\varpi_1$ and $\Omega_1$ still behaves as connection and curvature under the Lorentz gauge group $\SO$, as warranted by 
 $s_Lu_1$ in~\eqref{CC_1}. This is a hint, according to the example of GR treated in section~\ref{Application to the geometry of General Relativity}, that a new dressing operation to neutralize the Lorentz symmetry can be performed.

 \paragraph{Second reduction and final BRS algebra.}
 
$\!\!$With the dressing $u_0\!:\!\U\!\rarrow\!GL_{m+2}(\doubleR)\supset SO(1,m-1)$ given by~\eqref{u_0_0}, beside the composite fields $\varpi_0:=\varpi_1^{u_0}$ and
$\Omega_0:=\Omega_1^{u_0}$ respectively given by~\eqref{final varpi} and~\eqref{final Omega}, one construct the second composite ghost
 \begin{align*}
 \h v_0&=u_0\- \h v_1 u_0 + u_0\-s_0u_0 \\
&= u_0\-\big( u_1\-v_{\mathsf{W}} u_1 + u_1\-s_{\mathsf{W}}u_1 \big)u_0 + u_0\-v_Lu_0 + u_0\-s_{\mathsf{W}}u_0 + u_0\-s_Lu_0.
 \end{align*}
 The BRS transformations of $u_0$ under $s_{\mathsf{W}}$ and $s_L$ are respectively obtained by linearizing $u_0^W$ \eqref{u_0_W} and $u_0^S$ \eqref{u_0_0}:
  \begin{align}
s_{\mathsf{W}}u_0=\tilde \epsilon u_0 \quad \rarrow \quad \begin{pmatrix} 0 & 0 & 0 \\ 0 & s_{\mathsf{W}} e & 0 \\ 0 & 0 & 0 \end{pmatrix}=\begin{pmatrix} 0 & 0 & 0 \\ 0 & \epsilon & 0 \\ 0 & 0 & 0 \end{pmatrix}\begin{pmatrix} 1 & 0 & 0 \\ 0 & e & 0 \\ 0 & 0 & 1 \end{pmatrix}\qquad \text{and} \qquad s_Lu_0=-v_Lu_0
\label{u_0^L}
  \end{align}
The \emph{final composite ghost} is then,
\begin{align}
\h v_0&=u_0\-\big( u_1\-v_{\mathsf{W}} u_1 + u_1\-s_{\mathsf{W}}u_1 \big)u_0 + u_0\-v_Lu_0 + u_0\-\tilde \epsilon u_0 + u_0\-\big(-v_Lu_0\big),\notag \\
&=u_0\-\big( u_1\-v_{\mathsf{W}} u_1 + u_1\-s_{\mathsf{W}}u_1 \big)u_0 +\tilde \epsilon u_0\-u_0, \notag \\[1em]
&=\begin{pmatrix} 1 & 0 & 0 \\ 0 & e\- & 0 \\ 0 & 0 & 1 \end{pmatrix} \begin{pmatrix} \epsilon &  \d \epsilon\cdot e\- & 0 \\ 0 & 0 & \eta\-(e\-)^T\cdot \d\epsilon \\ 0 & 0 &-\epsilon \end{pmatrix}\begin{pmatrix} 1 & 0 & 0 \\ 0 & e & 0 \\ 0 & 0 & 1 \end{pmatrix}+ \begin{pmatrix} 0 & 0 & 0 \\ 0 & \epsilon \delta & 0 \\ 0 & 0 & 0 \end{pmatrix},\notag \\[1em]
&= \begin{pmatrix} \epsilon & \d\epsilon & 0 \\ 0 & \epsilon\delta  & g\- \d\epsilon \\ 0 & 0& -\epsilon \end{pmatrix}=: {\h v}_{\mathsf{W}}. 
\label{final-Weyl-ghost}
\end{align}
   Comparison with the first composite ghost \eqref{first-comp-ghost} shows that the ghost $v_L$ has been killed by the second dressing $u_0$,  and the Lorentz subsymmetry corresponding to $s_L$ is now trivial,
  \begin{align}
   s_L\varpi_0=0 \quad \text{and} \quad s_L\Omega_0=0. 
  \end{align} 
  Furthermore the dressing $u_0$ satisfies the compatibility condition $s_i u_0=0$, which is  the infinitesimal version of \eqref{u_0_1}, so that we also have
    \begin{align}
   s_i \varpi_0=0 \quad \text{and} \quad s_i \Omega_0=0.
  \end{align} 
    The triviality of these two subalgebras expresses the $\K_1$- and $\SO$-invariance of the composite fields $\varpi_0$ and $\Omega_0$. Their \emph{infinitesimal residual Weyl gauge transformations} are given by the \emph{final reduced BRS algebra}, 
with ${s_{\mathsf{W}}}^2=0$,
    \begin{align}
\label{Weyl-BRS-algebra}
    s_{\mathsf{W}}\varpi_0=-D_0\h v_{\mathsf{W}}, \qquad s_{\mathsf{W}}\Omega_0=\big[\Omega_0, \h v_{\mathsf{W}}\big],\qquad \text{and} \qquad s_{\mathsf{W}}\h v_{\mathsf{W}}= -{\h v_{\mathsf{W}}}^{\, 2},
    \end{align}
    with a final composite ghost $\h v_{\mathsf{W}}$ depending only on the Weyl ghost $\epsilon$ and its first order derivatives.
    
\paragraph{Two steps in one.}  
      
Mirroring the finite version given in section~\ref{2dressings}, we can reduce the inital BRS algebra with $s =s_{\mathsf{W}}+s_L+s_i$ to the final residual BRS algebra with $s_{\mathsf{W}}$ in a single step thanks to the dressing field $u=u_1u_0: \U \rarrow K_1\, GL_{m+2}(\doubleR)$. 
The corresponding composite ghost reads,
 \begin{align*}
 \h v &= u\- v u + u\- su
 = u\- ( v_{\mathsf{W}}+v_L +v_i )u + u\- ( s_{\mathsf{W}}+s_L +s_i )u
 \end{align*}
 One collects the BRS variations of the two dressing fields $u_1$ and $u_0$ 
 \begin{align*}
   s_i u_1=-v_i u_1,\qquad s_L u_0= - v_L u_0, \qquad  \text{and} \qquad  s_L u_1 = \big[u_1, v_L\big],  \qquad s_i u_0=0, 
 \end{align*}
the last two being \emph{compatibility conditions},\footnote{See \cite{JF_PhD} for a general treatment in the BRS case.} and proves $(s_L + s_i) u =  - (v_L +v_i)u$.
This means that $u$ is a dressing under the subgroup $\SO \, \K_1$. Hence, the composite ghost reduces to~\eqref{final-Weyl-ghost} since
\begin{align*}
\h v =u\-  v_{\mathsf{W}} u + u\-  \ s_{\mathsf{W}}u = \h v_{\mathsf{W}}.   
\end{align*}
 We see right away that $v_L$ and $v_i$ have been killed by the dressing $u$ so that the corresponding subalgebras are trivial,
\begin{align*}
s_i \varpi_0 =0 \quad \text{and}\quad s_i \Omega_0=0; \qquad
s_L\varpi_0=0 \quad \text{and}\quad s_L\Omega_0=0.
\end{align*}
The triviality of these two subalgebras means that the composite fields $\varpi_0$ and $\Omega_0$ display only a residual Weyl gauge freedom handled by the \emph{residual Weyl BRS algebra}  given by~\eqref{Weyl-BRS-algebra}. 

\medskip
To sum up, one has a dressing field 
\begin{equation}
u := u_1u_0 = \begin{pmatrix} 1 & q & \tfrac{1}{2}qq^t \\ 0 & \1 & q^t \\ 0 & 0 & 1 \end{pmatrix}\begin{pmatrix} 1 & 0 & 0 \\ 0 & e & 0 \\ 0 &  0 & 1 \end{pmatrix}
\end{equation}
with $u_1$ defined by $a-q\theta=0$ a gauge fixing like condition depending {\em locally} on entries of the gauge field~$\varpi$.
Upon writing
\[
s u\, u^{-1}= 
\begin{pmatrix} 
0 & -\iota & 0 \\ 0 & -v_L & -\iota^t \\ 0 & 0 & 0
\end{pmatrix}
+ 
\begin{pmatrix} 
0\ & \partial\epsilon\!\cdot\!e^{-1}& -\epsilon qq^t  \\
0 & \epsilon \1 &\ (\partial\epsilon\!\cdot\!e^{-1} )^t - 2\epsilon q^t\\
0 & 0 &0
\end{pmatrix} =: - \ell + \varrho u\-
\]
the BRS variation reads $su = - \ell u + \varrho$, with $\varrho= s_{\mathsf{W}} u$, see~\eqref{su_residuel}.
The first ghost matrix $-\ell = -v_L - v_i$ in the r.h.s. cancels out both the $\K_1$ and Lorentz $\SO(1,m-1)$ actions according to
\begin{align*}
\h v = u\- v u + u\- su = u\- (\ell + v_{\mathsf{W}}) u + u\- (- \ell u + \varrho) =  u\- v_{\mathsf{W}} u + u\- \varrho = {\h v}_{\mathsf W}
\end{align*}
as a realisation of formula~\eqref{final-ghost}.

  \paragraph{The residual Weyl BRS algebra: explicit results.}      
  
We restrict ourselves to the {\em normal} Cartan geometry. 
Let us recall  (see~\eqref{NCC} and~\eqref{NCC_curv})
 \begin{align*}
 \varpi_0:=\varpi^u =  \begin{pmatrix} 0 & P & 0 \\ dx & \Gamma & g\- \!\!\cdot\!P^T \\ 0 & dx^T\!\!\cdot\! g & 0 \end{pmatrix},\qquad \text{and} \qquad \Omega_0:=\Omega^u=\begin{pmatrix} 0 & C & 0 \\ 0 & W & g\-\!\!\cdot\!C^T \\ 0 & 0 & 0 \end{pmatrix}.
 \end{align*}
The residual BRS algebra given by~\eqref{Weyl-BRS-algebra} is straightforwardly computed thanks to the matrix form. The infinitesimal residual Weyl gauge transformation of the dressed \emph{normal} conformal Cartan connection is,
\begin{align*}
\MoveEqLeft
    s_{\mathsf{W}}\varpi_0 = -d\h v_{\mathsf{W}} - \varpi_0 \h v_{\mathsf{W}} -\h v_{\mathsf{W}}\varpi_0 \\[2mm]
   & = \begin{pmatrix} -d\epsilon &\ -d(\d\epsilon) &  0 \\ 0 & -d\epsilon \delta & -d(g\-  \!\cdot\!\d\epsilon) \\ 0 & 0 & d\epsilon  \end{pmatrix}
-
\begin{pmatrix} 0 & P \!\cdot\! \epsilon \delta & Pg\-\d\epsilon \\ dx \epsilon  &\ dx \d\epsilon   + \Gamma \epsilon \delta & \ \ \Gamma g\-\d\epsilon - g\- P^T\epsilon \\ 0 & dx^T\!\!\cdot\! g\, \epsilon \delta & dx^T\!\!\cdot\! g\,  g\- \d\epsilon\end{pmatrix} \\
    & \hskip 5cm -  
\begin{pmatrix} \d\epsilon dx  & \epsilon P + \d\epsilon \Gamma & \d\epsilon g\- P^T \\
\epsilon \delta dx &\ \ \epsilon\delta\Gamma +g\-d\epsilon dx^T\!\!\cdot\! g &\ \ \epsilon\delta g\-\!\cdot\! P^T \\
0 & -\epsilon dx^T\!\!\cdot\! g & 0  \end{pmatrix}.
\end{align*}
Reminding that $\epsilon$ anticommutes with forms of odd degree, that $d=dx\!\cdot\! \d$ and using the metricity condition $\nabla g\- = dg\-+g\-\Gamma^T +\Gamma g=0$ in the computation of entry (2, 3),  we obtain
\begin{multline*}
\begin{pmatrix} 0 & s_{\mathsf{W}}P & 0 \\
s_{\mathsf{W}}dx & s_{\mathsf{W}}\Gamma & s_{\mathsf{W}}\big(g\-\!\cdot\! P^T\big) \\
0 & s_{\mathsf{W}}\big(g\!\cdot\! dx\big) & 0 \end{pmatrix} = \\[2mm]
 \begin{pmatrix}  0 & - \nabla \d\epsilon
& 0 \\ 0 &\ -d\epsilon \delta - dx \d\epsilon - g\-d\epsilon dx^T\!\cdot\! g & \ \
-g\- \!\cdot\! (\nabla \d\epsilon)^T - 2\epsilon g\-\!\!\cdot\! P^T \\ 
0 & 2\epsilon dx^T\!\cdot\! g&0\end{pmatrix},
\end{multline*}
where, $\nabla \d\epsilon = d(\d\epsilon) + \d\epsilon\!\cdot\! \Gamma$, is the covariant derivative. 
  Let us detail each entry in components. Entry (2,1) is $s_{\mathsf{W}}dx^\mu =0$ and expresses the invariance of the coordinate chart. 
  This will be of constant use for the other entries. Entry (3, 2) is then,
  \begin{align}
  \label{s_w_metric}
  s^{}_{\mathsf{W}}\, g_{\mu\nu}  = 2\epsilon\, g_{\mu\nu}, 
  \end{align}
    which gives the infinitesimal  \emph{Weyl rescaling of the metric} tensor, see~\eqref{metric-W}. Entry (2, 2) reads in components,
    \begin{align}
    \label{s_w_gamma}
    s_{\mathsf{W}} {\Gamma^\rho}_{\mu\nu} = {\delta^\rho}_\nu \d_\mu\epsilon \ + \ {\delta^\rho}_\mu \d_\nu\epsilon \ + \  g^{\rho\lambda}\d_\lambda\epsilon g_{\mu\nu},
    \end{align}
 which is the infinitesimal \emph{Weyl transformation of the Christoffel symbols}, of the Levi-Civita connection.  Entry (1, 2) is,
 \begin{align}
 \label{s_w_schouten}
 s_{\mathsf{W}} P_{\mu\nu}=\d_\mu(\d_\nu \epsilon) - \d_\lambda\epsilon {\Gamma^\lambda}_{\mu\nu} = \nabla_\mu(\d_\nu\epsilon), 
 \end{align}
 which is the infinitesimal \emph{Weyl transformation of the Schouten tensor}. 
Finally entry (2, 3) is, 
 \begin{align}
 s_{\mathsf{W}}\big( g^{\rho\lambda}P_{\lambda\mu}\big) =  -2\epsilon g^{\rho\lambda}P_{\lambda\mu}+ g^{\rho\lambda}\bigg( \d_\mu(\d_\lambda\epsilon) - {\Gamma_{\mu\lambda}}^\alpha \d_\alpha \epsilon  \bigg) .
 \end{align}
    This is redundant with \eqref{s_w_metric} and \eqref{s_w_schouten}.  
    
    Comparing with the finite transformations given in section~\ref{Residual  Weyl  symmetry}  we see that the residual BRS algebra gives very easily the complete infinitesimal counterpart. Except for the Schouten tensor because the latter has a finite transformation which contains terms of order two in the Weyl parameter $\epsilon$. These terms are of course out of reach for the linear approximation captured by the BRS machinery. 
    
\smallskip
The infinitesimal Weyl gauge transformation of the dressed normal curvarture is given by
$s_{\mathsf{W}}\Omega_0= \Omega_0 \h v_{\mathsf{W}} -\h v_{\mathsf{W}}\Omega_0.$
Remembering this time that $\epsilon$ commutes with even forms and using $Wg\-=-g\-W^T$ (due to the $\eta$-skew symmetry of $F_1$), we obtain
 \begin{align*}
 s_{\mathsf{W}}\Omega_0=\begin{pmatrix} 0 & s_{\mathsf{W}}C & 0 \\ 0 &\ s_{\mathsf{W}} W &\ s_{\mathsf{W}}\big(g\- \!\cdot\! C^T\big)\\ 0 & 0 & 0 \end{pmatrix} = \begin{pmatrix} 0 &\ -\d\epsilon \!\cdot\! W & 0 \\ 0 & 0 & -g\-\! \!\cdot\!W^T  \!\cdot\! \d\epsilon^T -2\epsilon g\-  \!\cdot\! C^T\\ 0 &0 &0  \end{pmatrix}.
 \end{align*}
 Using again the fact that $s_{\mathsf{W}}dx^\mu=0$ we can write the entries in components.  Entry (1, 2) gives, 
 \begin{align}
 \label{s_w_cotton}
 s_{\mathsf{W}} C_{\nu, \mu\sigma} = -\d_\lambda\epsilon {W^\lambda}_{\nu, \mu\sigma},
 \end{align}
  which is the infinitesimal \emph{transformation of the Cotton tensor} under Weyl rescaling. Entry (2, 2) gives
 \begin{align}
 s_{\mathsf{W}} {W^\rho}_{\nu, \mu\sigma} = 0,
 \end{align}
 which states  the \emph{invariance of the Weyl tensor} under Weyl rescaling. Finally, entry (2, 3) is
  \begin{align}
  s_{\mathsf{W}}\big( g^{\rho\lambda}C_{\lambda, \mu\sigma} \big) = -2\epsilon g^{\rho\lambda}C_{\lambda, \mu\sigma} - g^{\rho\lambda}{{W_\lambda}^\alpha}_{\!\!, \mu\sigma} \d_\alpha\epsilon. 
  \end{align}
 This is redudant with \eqref{s_w_cotton} and \eqref{s_w_metric}.   
Once more, it should be pointed out how easily the modified BRS algebra provides the complete infinitesimal counterparts of the finite transformations derived in section~\ref{Residual  Weyl  symmetry}. 
 
\medskip
At last, the identity satisfied by the final composite ghost is, 
 \begin{align}
 s_{\mathsf{W}} \h v&= -\, {\h v_{\mathsf{W}}}^{\,2 }  
 \quad \rarrow \quad 
  \begin{pmatrix} s_{\mathsf{W}}\epsilon & s_{\mathsf{W}}(\d\epsilon) & 0 \\ 0 & s_{\mathsf{W}} \epsilon\delta  & s_{\mathsf{W}}\big( g\- \d\epsilon\big) \\ 0 & 0& -s_{\mathsf{W}}\epsilon \end{pmatrix} = \begin{pmatrix} 0 & 0 & 0 \\ 0 & 0 & -2\epsilon g\- \d\epsilon \\ 0 & 0 & 0  \end{pmatrix},
 \end{align}
 by recalling that $\epsilon$ anticommutes with itself (the same holds for $\d\epsilon$). 
 This just gives back the Weyl rescaling of the (inverse) metric $s_{\mathsf{W}}g\-=-2\epsilon g\-$ which is redundant with \eqref{s_w_metric}, but also
 \begin{align}
 s_{\mathsf{W}} \epsilon=0,
 \end{align}
 which expresses the fact that the residual Weyl gauge group  is abelian.

\bigskip
As a byproduct of this section, let us consider
the \emph{composite algebraic connection} introduced in Corollary~\ref{Modified Russian formula}. It reads,
 \begin{align}
\varpi_0 + \h v_{\mathsf{W}} = \begin{pmatrix} \epsilon &\ \ P +\d\epsilon  & 0 \\ dx &\ \ \Gamma + \epsilon\, \delta &\  \ g\- \!\!\cdot\!(P+ \d\epsilon)^T \\   0 &\ \  dx^T \!\!\cdot\! g & -\epsilon \end{pmatrix}.
\label{cac}
 \end{align}
Structurally, the algebraic connection is expressed as combinations between the metric $g$ and the Weyl ghost $\epsilon$ together with their derivatives. In~\cite{Boulanger2} these combinations have been obtained through a completely different approach and turn out to be relevant for the algebraic study of the Weyl anomaly.
Noteworthy, let us recall that in the present paper, we were able to use the well-tested BRS setting~\cite{Sto84,Stora:1985ac} on the normal conformal Cartan connection, and we have subsequently modified the differential algebra by the dressing field method. The route followed is rather robust and gives a clear well-grounded geometrical picture from which these combinations naturally emerge. 

\newpage
\section{Conclusion} 
\label{conclusion}

In this paper, with a clear geometric view, we exhibited an example that extends the dressing field method given in \cite{GaugeInv} to the second order conformal structure for which two dressing operations have been performed. Their composition was secured by the fact that the two dressing fields fulfilled compatibility conditions regarding their transformation laws with respect to various symmetry sectors. The scheme can be generalized to any number of dressing fields, for instance to higher order $G$-structures~\cite{JF_PhD}. 

In this example, treated in the very useful matrix notation, the final composite fields give the Riemannian parametrization of the normal conformal Cartan connection together with its curvature. The remaining symmetry has been shown to be the Weyl rescalings. The residual transformation of the composite fields provides at once well-known conformal transformations of noticeable tensors in conformal geometry of $\M$, namely, the (pseudo-Riemannian) metric, the Schouten, Cotton and Weyl tensors and the Christoffel symbols of the Levi-Civita connection. This is summarized in Table~\ref{normal-case}. In short, the dressing scheme provides gauge like transformation recombinations of the fields which amount to eliminating spurious degrees of freedom to the benefit  of geometrical objects together with their properties under the Weyl symmetry. 

Finally, the BRS counterpart of the dressing field method has been exhibited. Its central object is the \emph{composite ghost} which seems to correspond to a ``field dependent change of the generators'' of the differential algebra~\cite{Stora:2005tp} and encodes the residual gauge symmetry of the composite field, if any. Applied to the second order conformal structure, the composite algebraic connection (composite field $+$ composite ghost) is shown to give a structural geometric interpretation of the cohomological results obtained in~\cite{Boulanger2} regarding the Weyl symmetry. Morevover the corresponding \textit{modified BRS algrebra} provides in a very effective manner the linearized version of the residual gauge symmetry of Weyl rescalings derived in the firts part of the paper. 
A parallel can be made with the electroweak part of the standard model~\cite{Masson-Wallet,GaugeInv}. The erased subgroup, $SU(2)$ is the mirror of the Lorentz and special conformal transformations, while for the abelian residual gauge symmetry, $U(1)$ falls together with the Weyl dilations.

All the computations were performed at the classical level and were governed by geometrical considerations.
It deserves to see how all the process goes through the quantization, in particular the question of the quantum version of the composite fields and how to manage the erasure of gauge sub-symmetries.

Furthermore, one could push further ahead the use of the BRS techniques by combining together conformal gauge transformations and diffeomorphisms (of  $\M$).  This raises the question of the compatibility with the dressing field method in order to reduce the whole mixed symmetry to the Diff$\ltimes$Weyl symmetry. 
This issue will be addressed elsewhere~\cite{FLM} in a companion paper.
Moreover, the BRS differential algebra stemming from the second order conformal structure combined with the dressing method seems to offer an appropriate geometrical framework to tackle the Weyl anomalies with a standpoint grounded on the well-tested BRS approach. This matter is still under investigation.
  
\section*{Acknowledgments}

We would like to thank R.~Stora and T.~Schücker for fruitful discussions at the earlier stages of this work.

This work has been carried out in the framework of the Labex ARCHIMEDE (Grant No. ANR-11-LABX-0033)
and of the A*MIDEX project (Grant No. ANR-11-IDEX-0001-02), funded by the “Investissements d’Avenir”
French Government program managed by the French National Research Agency (ANR).

\appendix

\section{Appendix}
\label{Appendix}

For the sake of completeness, this appendix concerns the Weyl transformations (both finite and BRS versions) on  the first composite fields $\varpi_1$ and $\Omega_1$ obtained in~\eqref{varpi_1} and~\eqref{Omega_1} respectively. 

Their finite transformations are computed similarly to~\eqref{comp-field-W} upon using $u_1^W$ given in~\eqref{u_1^W}. A straightforward matrix computation yields, on the one hand,
\begin{align}
\label{varpi_1_W}
\varpi_1^W :\!&= 
 (\varpi^{u_1})^W = (\varpi^W)^{u_1^W}= (\varpi_1^W)^{W\-u_1k_1W} \notag\\
& = \varpi_1^{k_1W}= (k_1W)\- \varpi_1 (k_1W) + (k_1W)\-d(k_1W) \\[1mm]
 \begin{pmatrix} 0 & \alpha_1^W & 0  \\[1mm] \theta^W & A_1^W & {\alpha_1^t}^W \\[1mm] 0 &  {\theta^t}^W & 0  \end{pmatrix} & = \setlength{\arraycolsep}{3pt}
\begin{pmatrix}
0 & z\-\big( \alpha_1 + D(\zeta\!\cdot\! e\-) - (\zeta\!\cdot\! e\-) \theta (\zeta\!\cdot \!e\-)+ \tfrac{1}{2} (\zeta\!\cdot \!e\-) (\zeta\!\cdot \!e\-)^t \theta^t \big)  &  0  \\ 
z\theta &  A_1 + \theta(\zeta\!\cdot \!e\-) - (\zeta\!\cdot \!e\-)^t\theta^t & *
\\
0 & z\theta^t & 0 
\end{pmatrix} \notag
\end{align}
where $D(\zeta\cdot e\-) =d(\zeta\!\cdot e\-) - (\zeta\cdot\! e\-) A_1$ is the covariant derivative with respect to the spin connection $A_1$, where we have set $\zeta=z\-\d z$ as in the main text, and where $*= (\textrm{entry} (1,2) )^t$.

\noindent 
On the other hand,
\begin{align}
\label{Omega_1_W}
&\Omega_1^W:=   \begin{pmatrix}  f_1^W &\ \Pi_1^W & 0 \\ \Theta_1^W &\ F_1^W &\ {\Pi_1^t}^W \\ 0 &\ {\Theta_1^t}^W & -f_1^W \end{pmatrix}  
= (\Omega^W)^{u_1^W}=(\Omega_1^W)^{W\-u_1k_1W}=\Omega_1^{k_1W}=(k_1W)\- \Omega_1 k_1W \\[2mm]
& 
= \setlength{\arraycolsep}{3pt}\begin{pmatrix}  
f_1 \!-\! (\zeta\!\cdot\! e\-)\Theta  &   z\-\big(\Pi_1 \!-\! (\zeta\!\cdot\! e\-) (F_1 \!-\! f_1) \!-\! (\zeta\!\cdot\! e\-)\Theta(\zeta\!\cdot\! e\-) \!+\! \tfrac{1}{2}(\zeta\!\cdot\! e\-)(\zeta\!\cdot\! e\-)^t\Theta^t  \big) & 0  \\
z\Theta & F_1 \!+\!\Theta (\zeta\!\cdot\! e\-) \!-\! (\zeta\!\cdot\! e\-)^t\Theta^t & * \\
0 & z\Theta^t & * \end{pmatrix} \notag
\end{align}
where entry $(2,3)=(\mbox{entry }(1,2))^t$ and entry $(3,3)= -\, \mbox{entry }(1,1)$.

\medskip
In the {\em normal} case, see~\eqref{NCC_1}, formula~\eqref{varpi_1_W} remains formally unchanged but $\alpha_1$ becomes the Schouten $1$-form, and the Weyl variation~\eqref{Omega_1_W} reduces to (see~\eqref{Omega_1})
\begin{align}
\label{Omega_1_W_normal}
\Omega_1^W= \begin{pmatrix}  0  &\ \Pi_1^W & 0 \\ 0 &\ F_1^W &\ {\Pi_1^t}^W \\ 0 & 0 & 0 \end{pmatrix} = 
\setlength{\arraycolsep}{10pt}\begin{pmatrix}  
0  &   z\-\big( \Pi_1 - (\zeta\!\cdot\! e\-)F_1 \big)  & 0  \\
0  &  F_1 &  (\textrm{entry} (1,2))^t \\
0 & 0 & 0
\end{pmatrix},
\end{align}
with $\Pi_1=d\alpha_1 + \alpha_1 A_1$ is the Cotton $2$-form, and $F_1 = dA_1 + A_1^2 + \theta\alpha_1 - (\theta\alpha_1)^t$ is the Weyl $2$-form. 

\medskip
Let us now turn to the corresponding BRS setting. 
 It comes from the modification of the initial BRS algebra by the dressing field $u_1$. The matrix avatar of the Weyl ghost $\epsilon$ is given by setting $v_L\equiv 0$ in the ghost~\eqref{first-comp-ghost} and we keep the same notation $\h v_1$.
For the composite field $\varpi_1$, one has thus
\begin{align}
s_W \varpi_1 &= - d \h v_1 - \h v_1\varpi_1 - \varpi_1 \h v_1  \notag\\[2mm]
\setlength{\arraycolsep}{5pt}
 \begin{pmatrix} 0 & s_W\alpha_1  & 0  \\[1mm] s_W\theta & s_W A_1 & s_W\alpha_1^t \\[1mm] 
0 &  s_W \theta^t & 0  \end{pmatrix}
&= \setlength{\arraycolsep}{10pt}
\begin{pmatrix} 0 & - d(\d\epsilon\!\cdot\! e\-) - (\d\epsilon\!\cdot\! e\-) A_1 - \epsilon \alpha_1 & 0 \\
\epsilon \theta & - \theta (\d\epsilon\!\cdot\! e\-) - (\d\epsilon\!\cdot\! e\-)^t\theta^t & * \\
0 & \epsilon \theta^t & 0 \end{pmatrix}
\end{align}
where $*=(\mbox{entry }(1,2))^t$. For its curvature $\Omega_1$
\begin{align}
s_W \Omega_1 &=  \Omega_1 \h v_1 - \h v_1 \Omega_1   \notag\\[2mm]
\setlength{\arraycolsep}{5pt}
 \begin{pmatrix} 0 & s_W\Pi_1  & 0  \\[1mm] 0 & s_W F_1 & s_W\Pi_1^t \\[1mm] 
0 &  0 & 0  \end{pmatrix}
&= \setlength{\arraycolsep}{5pt}
\begin{pmatrix} 0 & - \epsilon \Pi_1 - (\d\epsilon\!\cdot\! e\-) F_1  & 0 \\
0 & 0 & * \\
0 & 0 & 0 \end{pmatrix}
\end{align}
~where $*=(\mbox{entry }(1,2))^t$.

\smallskip
These are the Weyl transformations of the normal conformal Cartan connection in the internal Minkowski indices before the second dressing by $u_0$. As seen in the main text, the dressing field $u_0$ allows to switch to spacetime (greek) indices.

\bibliographystyle{JHEP}

\bibliography{Biblio}
\addcontentsline{toc}{section}{References}

\end{document}